\documentclass[sn-mathphys,Numbered,pdflatex]{sn-jnl}
\usepackage{graphicx}%
\usepackage{multirow}%
\usepackage{amsmath,amssymb,amsfonts}%
\usepackage{amsthm}%
\usepackage{mathrsfs}%
\usepackage[title]{appendix}%
\usepackage{xcolor}%
\usepackage{textcomp}%
\usepackage{manyfoot}%
\usepackage{booktabs}%
\usepackage{algorithm}%
\usepackage{algorithmicx}%
\usepackage{algpseudocode}%
\usepackage{listings}%
\usepackage[utf8]{inputenc}
\usepackage{amsmath}
\usepackage{slashed}
\usepackage{graphicx}
\usepackage{amsfonts}
\usepackage{amssymb}
\usepackage{caption} 
\usepackage{subcaption}
\usepackage{fancyhdr} 
\usepackage{hyperref}
\usepackage{physics}
\usepackage{mathrsfs}
\usepackage{xcolor}
\begin{document}
\title{Two dark matter candidates in a doublet-triplet Higgs model}
\author*[1,2]{\fnm{S.} \sur{Melara-Duron}} \email{sheryl.melara@unah.edu.hn}

\author[3]{\fnm{R.} \sur{Gaitán}}\email{rgaitan@unam.mx}

\author[3]{\fnm{J. M.} \sur{Lamprea}}\email{jmlamprea@gmail.com}

\affil*[1]{\orgdiv{Instituto de Física}, \orgname{Universidad Nacional Autónoma de México}, \orgaddress{\street{A.P. 20-364}, \city{Ciudad de México}, \postcode{01000}, \country{México}}}

\affil*[2]{\orgdiv{Departamento de Astronomía y Astrofísica, Facultad de Ciencias Espaciales}, \orgname{Universidad Nacional Autónoma de Honduras}, \orgaddress{\street{Bulevar Suyapa}, \city{Tegucigalpa}, \state{M.D.C.}, \country{Honduras}}}

\affil[3]{\orgdiv{Departamento de Física, Facultad de Estudios Superiores Cuautitlán}, \orgname{Universidad Nacional Autónoma de México}, \orgaddress{\city{Estado de México}, \postcode{54770}, \country{México}}}
\abstract{
 We study a Standard Model extension that provides a bi-component dark matter scenario as well as a mechanism for the generation of Left-Handed neutrino masses. We extend the Standard Model scalar sector by adding an inert $SU(2)_L$ complex doublet with hypercharge $Y= 1/2$ and a real inert $SU(2)_L$ triplet with hypercharge $Y=0$. These scalars provide suitable dark matter candidates belonging to two different dark sectors that are stabilised by the discrete symmetries $Z_2 \times Z_2'$. 
 We consider the contribution of both candidates to the total relic abundance in order to reduce the desert regions in the Inert Doublet and Inert Triplet Models. In addition, we have added an active real scalar $SU(2)_L$ triplet with hypercharge $Y=1$ in order to generate light neutrino Majorana masses through the type-II seesaw mechanism. We have analysed the dark matter phenomenology for the model and alongside with the neutrino mass generation.
}

\keywords{scalar dark matter, beyond Standard Model, WIMPs, multi-Higgs models}

\maketitle

\section{Introduction}

The Standard Model (SM) of particle physics successfully explains most of the particle phenomena we observe in Nature~\cite{SGlashow1961, Salam1964, Weinberg1967}.
However, the SM cannot provide an explanation for some anomalies found in the Universe. Among these puzzles are the origin of light neutrino masses~\cite{SKamiokande98}, the mechanism behind the origin of the matter-antimatter asymmetry, and the existence of dark matter (DM)~\cite{Vera1970}. 
In spite of the fact that is has not been possible yet to detect DM, direct detection experiments impose upper bounds on the cross-section for DM-nucleon interactions. 
The current most stringent limits has been set by the PANDAX-II~\cite{PANDA2017}, LUX-ZEPLIN~\cite{LZ2023}, and XENONnT~\cite{XENONnT2023} experiments. 
DM candidates must satisfy those experimental constraints as well as theoretical ones~\cite{Taoso:2007qk}.

In order to describe DM and its interactions with ordinary matter, we need to go beyond the Standard Model (BSM).
The most studied BSM scenario includes particles like WIMPs (Weakly Interactive Massive Particles) as DM candidates. 
One of the most popular of those extensions is the Inert Doublet Model (IDM)~\cite{IDM2010, Krawczyk2015, Gaitan2020, Kalinowski_2021}, which consists of adding to the SM content an inert scalar $SU(2)_L$ doublet. This model provides a good DM candidate for the mass region $M_{DM} \geq 550$~GeV. 
However, for smaller mass values, $M_W < M_{DM} < 550$~GeV, the relic density of the DM turns to be under-abundant\footnote{This  is often referred as the desert region.} due to the large annihilation cross-section of dark matter in that region.
Another successful model is the Scotogenic model proposed by Ernest Ma~\cite{Ma2006}. 
This model, in addition to providing a good dark matter candidate as the IDM, also provides a mechanism to generate light neutrino masses at the radiative level. 
However, it also has the same under-abundant region as the IDM. 
In addition to the Inert Doublet and Scotogenic models, other scalar DM models have been proposed, such as singlet and triplet extensions~\cite{Fischer2011}. Here we are interested in Inert Triplet Models (ITM)~\cite{Akari2011, Khan:2016sxm} which postulate a scalar DM candidate as well, though a wider desert region: $M < 1970$~GeV~\cite{Jangid2020} is obtained compared to the IDM.  

A possible solution to the problem of under-abundance in the IDM and ITM is considering the contribution of two DM candidates to account for the total relic density. Two-component DM scenarios have recently become a topic of interest~\cite{Chakrabarty2022, Betancur_2022}. In this work, we study an extension of the SM with two dark sectors. We consider the neutral CP even component of a $SU(2)_L$ scalar doublet and the neutral component of a $SU(2)_L$ scalar triplet as our DM candidates. Provided that some DM mass regions are forbidden in each case separately, we aim to study DM-DM conversion between the two candidates to reduce some of those regions. Additionally to the study of the DM phenomenology, we have included an active triplet $\Delta$ with hypercharge $Y=1$ in order to generate  light neutrino Majorana masses through the type-II seesaw mechanism. It is worth mentioning that models with triplets have received attention since it is possible to generate baryogenesis via leptogenesis through the decay of heavy particles. 

This paper is organised as follows. In the first section~\ref{sec:Model}, we introduced our two dark sector model where dark matter candidates are both scalars. In Section~\ref{Sec:NeutrinoMasses}, we present neutrino mass generation through the type-II seesaw mechanism. Section~\ref{Sec:Constraints} discusses the relevant constraints in our model. Our results for DM phenomenology are presented and discussed in Section~\ref{Sec:Results}. Finally, we present our conclusions in Section~\ref{Sec:Conclusions}.

\section{The Model}\label{sec:Model}

\begin{table}[htb]
\centering
\begin{tabular}{|c|c|c|c|c|}
\hline
Particle  & $SU(2)_L$ & $U(1)_Y$ & $\mathbb{Z}_{2}$ & $\mathbb{Z}'_{2}$ \\
\hline
\hline
$\Phi_1$ & $2$ & $1/2$ & + & +\\
$\Phi_{2}$  & $2$ & $1/2$ & - & + \\
$\Delta$  & $3$ & 1 & + & +\\
$T$  & $3$ & 0 & + & -\\
\hline
\end{tabular}
\caption{Quantum numbers of the SM Higgs ($\Phi_1$) and additional scalar fields considered in the model.}\label{tab:1}
\end{table}

In the present study, we have considered an extension of the SM where we have included a complex scalar $SU(2)_L$ doublet $\Phi_2$ with $Y=1/2$, a real scalar $SU(2)_L$ triplet $T$ with hypercharge $Y = 0$ and a real scalar $SU(2)_L$ triplet $\Delta$ with hypercharge $Y = 1$. 
Our dark matter candidates come from the lightest neutral CP-even components of doublet $\Phi_{2}$ and triplet $T$. 
The stability of the DM candidate is guaranteed by the direct product of discrete $\mathbb{Z}_{2}$ symmetries $\mathbb{Z}_{2}\times \mathbb{Z}'_{2}$.  All SM fields are trivially charged under this symmetry, while $\Phi_{2}$ and $T$ belong to each dark sector $\mathbb{Z}_{2}$ and $\mathbb{Z}_{2}'$ respectively.
The relevant particle content and quantum numbers of all the added fields in the model are shown in Table~\ref{tab:1}. 

The most general scalar potential renormalisable and invariant under the SM gauge and $\mathbb{Z}_{2}\times \mathbb{Z}'_{2}$ symmetry can be written as:
\begin{equation}
    V = V_{\Phi_1 \Phi_2} + V_{\Delta} + V_{T} + V_{\text{int}},
\end{equation}
where,
\begin{align}
    V_{\Phi_1 \Phi_2}  &=  \mu_{1}^{2}\Phi_{1}^{\dagger}\Phi_{1} + \lambda_{1}(\Phi_{1}^{\dagger}\Phi_{1})^{2} + \mu_{2}^{2}\Phi_{2}^{\dagger}\Phi_{2} + \lambda_{2}(\Phi_{2}^{\dagger}\Phi_{2})^{2} + \lambda_{3}\Phi_{1}^{\dagger}\Phi_{1}\Phi_{2}^{\dagger}\Phi_{2} \notag
    \\
    &+\lambda_{4}\Phi_{1}^{\dagger}\Phi_{2}\Phi_{2}^{\dagger}\Phi_{1} 
    + \dfrac{1}{2}\lambda_{5}\left[(\Phi_{1}^{\dagger}\Phi_{2})^{2} + (\Phi_{2}^{\dagger}\Phi_{1})^{2}\right],\\
    V_{\Delta} &= \mu_{\Delta}^{2}\Delta ^{\dagger}\Delta + \lambda_{\Delta}(\Delta ^{\dagger}\Delta)^{2},\\
    V_{T} &=  \mu_{T}^{2}T^{\dagger}T + \lambda_{T}(T^{\dagger}T)^{2},\\
    V_{\text{int}} &= \kappa_{\Phi_{1}\Delta}\left(\Phi_{1}^{\dagger}\Delta\tilde{\Phi}_{1} + h.c.\right) + \kappa_{\Phi_{2}\Delta}\left(\Phi_{2}^{\dagger}\Delta\tilde{\Phi}_{2} + h.c.\right) \notag
    \\ 
    &+ \lambda_{\Phi_{1}\Delta}\Phi_{1}^{\dagger}\Phi_{1}\Delta^{\dagger}\Delta 
    + \lambda_{\Phi_{2}\Delta}\Phi_{2}^{\dagger}\Phi_{2}\Delta^{\dagger}\Delta + \lambda_{\Phi_{1}T}\Phi_{1}^{\dagger}\Phi_{1}T^{\dagger}T + \lambda_{\Phi_{2}T}\Phi_{2}^{\dagger}\Phi_{2}T^{\dagger}T.
\end{align}

We can ignore the mixing term between the SM Higgs $h$ and the neutral component of the active triplet $\Delta^0$, which is proportional to the terms $\kappa_{\Phi_{1}\Delta}v_{\Delta}$ and $v_{\Delta}^{2}$, since $v_{\Delta} \sim \mathcal{O}$~(1 eV) is small (in order to generate light neutrino masses\footnote{ See Section~\ref{Sec:NeutrinoMasses}.}) in comparison with the squared mass terms: $\mathcal{O} (m_h^2) \sim v^2 \sim 10^4$~GeV and $\mathcal{O} (m^2_{\Delta^0}) \sim \kappa_{\Phi_{1}\Delta} \, v^{2}/v_{\Delta} \sim $ 1 TeV. 
As the field $\Phi_1$ will be identified as the SM Higgs doublet, $\mu_{1}^{2}< 0$ is required by the Electroweak Symmetry Breaking (EWSB) in the SM. We have considered all the parameters in the scalar potential to be real.

The scalar fields can be parameterised as follows
\begin{align}
\Phi_{1} &= 
\begin{pmatrix}
0  \\ \dfrac{v + h}{\sqrt{2}}
\end{pmatrix},
\hspace{2.5cm}
\Phi_{2} =
\begin{pmatrix}
H^{+} \\
\dfrac{H^{0}+ iA^{0}}{\sqrt{2}}
\end{pmatrix},\\ 
\Delta &= 
\begin{pmatrix}
\dfrac{\Delta^{+}}{\sqrt{2}} & \Delta^{++}  \\ 
v_{\Delta} + \Delta^{0} & -\dfrac{\Delta^{+}}{\sqrt{2}}
\end{pmatrix},
\hspace{1.0cm}
T =
\begin{pmatrix}
\dfrac{T^{0}}{\sqrt{2}} & -T^{+} \\
-T^{-} & -\dfrac{T^{0}}{\sqrt{2}} 
\end{pmatrix}.
\end{align}
%
%

The Vacuum Expectation Values (VEV) of the fields $\Phi_1$ and $\Delta$ are labelled as $v$ and $v_{\Delta}$ respectively. The fields $\Phi_{2}$ and $T$ do not pick up a VEV as the discrete $\mathbb{Z}_2 \times \mathbb{Z}'_2$ is not broken. Therefore, $\Phi_2$ and $T$ do not couple to the SM fermions, i.e. are inert scalars. 

The masses of the physical scalars after EWSB are:
\begin{align}
m^{2}_{h} &= 2\lambda_{1}v^{2}, \\ 
m^{2}_{H^{\pm}} &= \mu^{2}_{2} + \dfrac{1}{2}\lambda_{3}v^{2} + \lambda_{\Phi_{2}\Delta}v^{2}_{\Delta},\\
m^{2}_{H^{0}} &= \mu^{2}_{2} + \dfrac{1}{2} \left(\lambda_{3}+\lambda_{4}+\lambda_{5}\right) v^{2}  + \kappa_{\Phi_{2}\Delta}v_{\Delta} + \lambda_{\Phi_{2}\Delta}v^{2}_{\Delta},\\
m^{2}_{A^{0}} &= \mu^{2}_{2} + \dfrac{1}{2}\left(\lambda_{3}+\lambda_{4} - \lambda_{5}\right)v^{2}  - \kappa_{\Phi_{2}\Delta}v_{\Delta} + \lambda_{\Phi_{2}\Delta}v^{2}_{\Delta},\\
m^{2}_{T^{\pm}, T^{0}} &= \mu^{2}_{T} + \dfrac{1}{2}\lambda_{\Phi_{1}T}v^{2},\\
m^{2}_{\Delta^{\pm \pm}} &= -\dfrac{1}{2}\kappa_{\Phi_{1}\Delta} \dfrac{v^{2}}{v_{\Delta}},\\
m^{2}_{\Delta^{\pm}, \Delta^{0}} &=-\dfrac{1}{4}\kappa_{\Phi_{1}\Delta} \dfrac{v^{2}}{v_{\Delta}},
\end{align}
where $\kappa_{\Phi_{1}\Delta}< 0$ and $v=246$ GeV. We identify $m_h = 125.09$ GeV~\cite{CMS2012,ATLAS2012} as the mass of the SM Higgs boson.

\begin{table}[h]
    \centering
    \begin{tabular}{|c|c|c|c|c|c|c|}
    \hline
    Particle  & $Q_{EM}$ &  $U(1)_{Y}$ & $SU(2)_{L}$ & $SU(3)_{C}$ & $\mathbb{Z}_2$ & $\mathbb{Z}_2'$ \\
    \hline
    \hline
     $h$   & 0 & 1/2 & 2 & 1 & + & +\\
     $H^{\pm}$ & $\pm 1$ & 1/2 & 2 & 1 & - & + \\
     $H^{0}$ & 0 & 1/2 & 2 & 1 & - & +\\
     $A^{0}$ & 0 & 1/2 & 2 & 1 & - & +\\
     $T^{\pm}$ & $\pm 1$ & 0 & 3 & 1 & + & -\\
     $T^{0}$ & 0 & 0 & 3 & 1 & + & - \\
     $\Delta^{\pm \pm}$ & $\pm 2$ & 1 & 3 & 1 & + & + \\
     $\Delta^{\pm}$ & $\pm 1$ & 1 & 3 & 1 & + & + \\
     $\Delta^{0}$ & 0 & 1 & 3 & 1 & + & +\\
     \hline
    \end{tabular}
    \caption{Quantum numbers of the physical scalars  after EWSB.}
    \label{tab:particles}
\end{table}

In table~\ref{tab:particles} we show the complete list of particles after EWSB and their corresponding quantum numbers. The lightest neutral $\mathbb{Z}_{2}$-odd particle is a DM candidate. Depending on our choice of the sign of $\lambda_{5}$, either $H_{0}$ or $A_{0}$ can be the lightest one. We have chosen $\lambda_{5} < 0$ so that the $H_{0}$ is our DM candidate. The second DM candidate is the lightest $\mathbb{Z}_{2}'$-odd particle, $T_{0}$. 

We observe that the mass splitting between the charged Higgs $H^{\pm}$ and $H_{0}$ , $m_{H^{\pm}}-m_{H^{0}}$, as well as the mass splitting between $A_{0}$ and $H_{0}$, $m_{A^{0}}-m_{H^{0}}$, are free parameters in the model. On the contrary, the charged and neutral components of the inert triplet $T$ have degenerate masses at tree level. However, due to radiative effects, there is a mass splitting at the one-loop level~\cite{Cirelli2020}
\begin{equation}
    \Delta m_{T} = \dfrac{\alpha}{2}M_{W}\sin^{2}\dfrac{\theta_{W}}{2},
\end{equation}
where $\alpha$ is the fine structure constant, $M_{W}$ the mass of the $W$ boson, $M_{Z}$ the mass of the Z boson and $\theta_{W} = \cos^{-1}(M_{W}/M_{Z})$ the weak mixing angle. Then,
\begin{equation}
    \Delta m_{T} = m_{T^{\pm}} - m_{T^{0}} = 166 ~\textnormal{MeV}.
    \label{eq:tripletsplitting}
\end{equation}

The free BSM parameters of the model are 16:\{$m_{H^{0}}$, $m_{A^{0}}$, $m_{H^{\pm}}$, $m_{T^{0}}$, $m_{T^{\pm}}$, $\lambda_{L}$, $\lambda_{\Phi_{1}T}$, $\lambda_{\Phi_{2}T}$, $\lambda_{2}$, $\lambda_{T}$, $\lambda_{\Delta}$, $\lambda_{\Phi_{1}\Delta}$, $\lambda_{\Phi_{2}\Delta}$, $m_{\Delta^{\pm\pm}}$, $\kappa_{\Phi_{2}\Delta}$, $v_{\Delta}$\},
where we have defined 
\begin{equation}
    \lambda_{L} = \dfrac{1}{2}\left( \lambda_{3} + \lambda_{4} + \lambda_{5}  \right),
\end{equation}
as an independent parameter. However, the relevant parameters in our model for relic density calculation are five masses $m_{H^{0}}$, $m_{H^{\pm}}$, $m_{A^{0}}$, $m_{T^{0}}$, $m_{T^{\pm}}$, and three quartic couplings $\lambda_{L}$, $\lambda_{\Phi_{1}T}$, $\lambda_{\Phi_{2}T}$.
We will restrict the value of these parameters in section~\ref{Sec:Constraints} using the respective theoretical and experimental constraints. 

\section{Neutrino Masses}\label{Sec:NeutrinoMasses}

The Yukawa part of the Lagrangian involved in the Left--Handed (LH) neutrino mass generation is given by
\begin{equation}
    \mathscr{L}_{Y} = Y_\Delta^{\alpha \beta} L^{T}_{\alpha} \,C \, i \tau_{2}\, \Delta \, L_{\beta} + \textit{h.c.},
\end{equation}
where $\alpha, \beta$ correspond to flavour indices, $C$ is the charge conjugation operator, $\tau_2$ is the second Pauli matrix, $\tilde{\Phi}_{2} = i \tau_2 \Phi_{2}^*$ and $L_\alpha = (\nu_\alpha, \ell_{L\alpha})$ is the SM LH lepton doublet with flavour $\alpha$. 

Due to the presence of the active triplet $\Delta$, the LH neutrino masses are generated through the type-II seesaw mechanism~\cite{MOHAPATRA2005257, Saiyad2022}:
\begin{equation}\label{eq:seesaw}
    m_{\nu}^{II} = \kappa_{\Phi_{1}\Delta}\dfrac{v^{2}}{M_{\Delta}^{2}}Y_{\Delta},
\end{equation}
where $Y_{\Delta}$ is the Yukawa coupling matrix of the triplet $\Delta$ with leptons. The VEV of the triplet is induced by the EWSB as, $v_{\Delta}\simeq \kappa_{\Phi_{1}\Delta}v^{2}/2M_{\Delta}^{2}$~\cite{MAGG198061, Valle1980, MohapatraSenja1981,WETTERICH1981343}. Therefore, we obtain:

\begin{equation}
   m_{\nu}^{II} = 2 v_{\Delta} Y_{\Delta}.
\end{equation}

According to the type-II seesaw equation~(\ref{eq:seesaw}), the mass of the scalar triplet $\Delta$ must be large enough to generate the correct left-handed neutrino masses. In addition, the VEV of the triplet must be very small~\cite{Datta2022}. For that reason we have chosen $M_{\Delta} = 1$ TeV and $v_{\Delta} = 1$ eV.

We can adjust the Yukawa coupling matrix $Y_{\Delta}$ in $m_{\nu}^{II}$ to produce the light neutrino masses according to neutrino oscillation data. Hence, we consider:
\begin{equation} \label{eq:21}
    m_{\nu}^{II} = U^{*}m_{\nu}^{d}U^{\dagger},
\end{equation}
where $m_{\nu}^{d} = \text{diag}(m_1, m_2, m_3)$ and $U$ is the Pontecorvo--Maki--Nakagawa--Sakata (PMNS) neutrino mixing matrix, parametrised as 
\begin{equation}
    U = 
    \begin{pmatrix}
    c_{12}\hspace{0.1cm}c_{13} & s_{12}\hspace{0.1cm}c_{13} & s_{13}\hspace{0.1cm}e^{-i\delta_{CP}} \\
    -s_{12}\hspace{0.1cm}c_{23} - c_{12}\hspace{0.1cm}s_{13}\hspace{0.1cm}s_{23}\hspace{0.1cm}e^{i\delta_{CP}} & c_{12}\hspace{0.1cm}c_{23} - s_{12}\hspace{0.1cm}s_{13}\hspace{0.1cm}s_{23}\hspace{0.1cm}e^{i\delta_{CP}} & c_{13}\hspace{0.1cm}s_{13} \\
    s_{12}\hspace{0.1cm}s_{13} - c_{12}\hspace{0.1cm}s_{13}\hspace{0.1cm}c_{23}\hspace{0.1cm}e^{i\delta_{CP}} & -c_{12}\hspace{0.1cm}s_{23} - s_{12}\hspace{0.1cm}s_{13}\hspace{0.1cm}c_{23}e^{i\delta_{CP}} & c_{13}c_{23}
    \end{pmatrix},
\end{equation}
where $c_{ij}\equiv \cos\theta_{ij}$, $s_{ij}\equiv \sin\theta_{ij}$ and $\delta_{CP}$ is the Dirac CP violation phase. 
We use the best-fit values of the neutrino mixing angles, CP phase and $\Delta m_{21}^{2}$ and $\Delta m_{32}^{2}$ as input parameters according to~\cite{PDG2022}. For simplicity, we consider $m_{1}=0$ in the case of Normal mass Ordering (NO) and $m_{3}=0$ for Inverted mass Ordering (IO), as well as the Majorana phases to be zero. For the NO we obtain the following Yukawa coupling matrix
\begin{equation}
    Y_{\Delta} = 10^{-3} \times
    \begin{pmatrix}
    1.361 + 0.560i  & -0.646 - 1.956i & -3.208 - 1.742i\\
    -0.646 - 1.956 i& 15.221 - 0.214 i & 10.697 + 0.008 i\\
    -3.208 - 1.742 i& 10.697 + 0.008 i& 12.300 + 0.185 i
    \end{pmatrix}.
\end{equation}
       
       
       
       
       
       
       
       
In the same way, for the IO we obtain
\begin{equation}
    Y_{\Delta} = 10^{-3} \times
    \begin{pmatrix}
    24.228  & -0.603 - 2.671 i & -0.758 - 2.354i \\
    -0.603 - 2.671 i & 10.623 + 0.134 i& -12.607 + 0.142 i\\
    -0.758 - 2.354 i & -12.607 + 0.142 i& 13.822 + 0.147 i
    \end{pmatrix}.
\end{equation}

Although the type-II seesaw mechanism is not related to the DM phenomenology, we have shown the light neutrino masses generation to give a more complete description of the model. Finally, we want to point out that for this seesaw mass generation is possible to generate leptogenesis which will be considered in a further study.

\section{Constraints}\label{Sec:Constraints}

In the following, we discuss the theoretical and experimental constraints imposed on the numerical scan.

\subsection{Experimental Constraints}\label{Sec:ExpConstraints}

\subsubsection{Dark Matter Abundance}

In order to provide appropriate DM candidates, our model must be able to yield the right DM abundance according to cosmological observations. From the latest Planck Collaboration~\cite{PLANCK2018} data analysis, the DM relic abundance should be within the range
\begin{equation}
    \Omega_{DM} h^{2} = 0.1200 \pm 0.0010.
\end{equation}

We have imposed the restriction that our results for the total relic density must lie within that band. 
In addition, we have included the most updated upper limits for the Spin-Independent (SI) cross-section from direct detection experiments. We used the limits set by PANDAX-II~\cite{PANDA2017}, LUX-ZEPLIN~\cite{LZ2023} and XENONnT~\cite{XENONnT2023} which are currently the most stringent limits.

\subsubsection{Constraints from Colliders}\label{subsec:colliderconst}

The absence of a signal within searches of supersymmetric neutralinos can be used to constrain the IDM in the LEP II~\cite{LEPII09, Krawczyk2013}; data analysis excludes the mass regions:
\begin{equation}
    m_{H^{0}} < 80 \hspace{0.1cm} \textnormal{GeV}, \hspace{0.5cm} m_{A^{0}} < 100 \hspace{0.1cm} \textnormal{GeV} \hspace{0.5cm} \textnormal{and} \hspace{0.2cm} \Delta m = m_{A^{0}}-m_{H^{0}} > 8 \hspace{0.1cm} \textnormal{GeV}. \label{eq:constraintsneu}
\end{equation}

The search of charginos from OPAL~\cite{OPAL2004} collaboration imposes the following bounds for the charged Higgs masses~\cite{Thaler_2007}:

\begin{equation}
    m_{H^{\pm}} \gtrsim 70 - 90 \hspace{0.1cm} \textnormal{GeV}.
    \label{eq:constraintschar}
\end{equation}
As well, there are recast studies of the LHC searches for invisible Higgs decays within Vector Boson Fusion~\cite{Dercks2019} and searches in final states with two leptons plus missing transverse energy~\cite{Belanger2015, Ilnicka2016}, which provide constraints on the IDM parameter space.  
\newline
Additionally, the LHC~\cite{CMS2019} sets constraints on the mass of the active scalar triplet $\Delta$ components:
\begin{equation}
    m_{\Delta^{\pm \pm}} > 820\hspace{0.1cm} \textnormal{GeV} \hspace{0.1cm} \textnormal{and} \hspace{0.1cm} m_{\Delta^{\pm}}> 350\hspace{0.1cm} \textnormal{GeV}.
\end{equation}

It is to be noted that we have used these constraints as a guide in the analysis in our model. They are only used to provide a first approximation to the allowed mass regions for the inert doublet and triplet scalars. A detailed collider study may provide different limits, which is beyond the scope of our work.

\subsection{Theoretical Constraints}\label{Sec:TheoCon}

\subsubsection{Positivity and stability of the scalar potential}

To ensure the potential remains stable and bounded from below, we need to determine the necessary vacuum stability conditions. Applying co-positivity criteria~\cite{Kannike2020}, we have obtained for the quartic couplings
\begin{equation}
\lambda_{1}>0, \hspace{0.5cm} \lambda_{2}> 0, \hspace{0.5cm}  \lambda_{\Delta} > 0, \hspace{0.5cm}  \lambda_{T}>0,
\end{equation}
and 
\begin{align}
\lambda_{3} + \lambda_{4} -\abs{\lambda_{5}} + \sqrt{\lambda_{1}\lambda_{2}} >0, \hspace{0.25cm} \lambda_{3} + \sqrt{\lambda_{1}\lambda_{2}} >0,\\
\lambda_{\Phi_{1}\Delta} + \sqrt{\lambda_{1}\lambda_{4}} >0, \hspace{0.25cm} \lambda_{\Phi_{2}\Delta} + \sqrt{\lambda_{2}\lambda_{\Delta}} >0, \\
\lambda_{\Phi_{1}T} + \sqrt{\lambda_{1}\lambda_{T}} >0, \hspace{0.25cm} \lambda_{\Phi_{2}T} + \sqrt{\lambda_{2}\lambda_{T}} > 0. 
\end{align}

\subsubsection{Unitarity}

We consider the restrictions coming from the unitarity of the scattering matrix~\cite{Ginzburg2005}. Unitarity demands the absolute eigenvalues of the scattering matrix to be less than $8\pi$. Imposing that constraint in our model we obtain,
\begin{equation*}
\abs{\lambda_{3} \pm \lambda_{4}} \leq 8\pi,~~\abs{\lambda_{3} \pm \lambda_{5}} \leq 8\pi,~~\abs{\lambda_{3} + 2\lambda_{4} \pm \lambda_{5}} \leq 8\pi,
\end{equation*}
\begin{equation}
\abs{\lambda_{T}} \leq 24\pi, \hspace{0.5cm} \abs{\lambda_{\Delta}} \leq 24\pi, \hspace{0.5cm} \abs{\lambda_{\Phi_{2}T}} \leq 8\pi,
\end{equation}
\begin{equation*}
    \abs{\lambda_{\Phi_{1}T}} \leq 8\pi, \hspace{0.5cm} \abs{\lambda_{\Phi_{1}\Delta}} \leq 8\pi,
    \hspace{0.5cm} \abs{\lambda_{\Phi_{2}\Delta}} \leq 8\pi
\end{equation*}
and
\begin{equation*}
\abs{\lambda_{1} + \lambda_{2} \pm \sqrt{(\lambda_{1}-\lambda_{2})^{2} + \lambda_{4}^{2}}} \leq 8\pi, \hspace{0.5cm} \abs{\lambda_{1} + \lambda_{2} \pm \sqrt{(\lambda_{1}-\lambda_{2})^{2} + \lambda_{5}^{2}}} \leq 8\pi.
\end{equation*}

\subsubsection{Perturbativity of the scalar potential}

The perturbativity of the scalar potential requires the parameters of the model must satisfy 
\begin{equation}
    \abs{\lambda_{i}} \leq 4\pi, \hspace{1.0cm} \abs{g_{i}}, \hspace{0.1cm} \abs{y_{\alpha\beta}} \leq \sqrt{4\pi},
\end{equation}
where $\lambda_{i}$ are the quartic couplings, while $g_{i}$ and $y_{\alpha\beta}$ denote the SM gauge and Yukawa couplings respectively.

\subsubsection{Oblique parameters}

Adding new fields to the Standard Model Lagrangian induces a contribution to the oblique parameters $S, T, U$~\cite{Peskin92}. We can write this contribution for $X = S, T, U$ as
\begin{equation}
\Delta X = \Delta X_{ID} + \Delta X_{IT} + \Delta X_{AT},
\end{equation}
where the subscripts ID, IT and AT denote the contribution from the inert doublet $H$, inert triplet $T$ and active triplet $\Delta$ respectively. 

In the case of the inert doublet, we obtain~\cite{GRIMUS200881}
\begin{align}
\begin{split}
\Delta S_{ID} &= \dfrac{1}{2\pi}\left[\dfrac{1}{6} \log\left(\dfrac{m^{2}_{H^{0}}}  {m^{2}_{H^{\pm}}}\right) - \dfrac{5}{36} + \dfrac{m^{2}_{H^{0}} m^{2}_{A^{0}}}{3(m^{2}_{A^{0}}-m^{2}_{H^{0}})^{2}}\right] \\
&+ \dfrac{1}{2\pi}\left[\dfrac{m^{4}_{A^{0}}(m^{2}_{A^{0}}-3m^{2}_{H^{0}})}{6(m^{2}_{A_{0}}-m^{2}_{H^{0}})^{3}} \log \left(\dfrac{m^{2}_{A^{0}}}{m^{2}_{H^{0}}}\right)\right],
\end{split}
\\
\Delta T_{ID} &= \dfrac{1}{16\pi s^{2}_{W} M^{2}_{W}}\left[F(m^{2}_{{H}^{+}}, m^{2}_{H^{0}})+ F(m^{2}_{H^{+}}, m^{2}_{A^{0}})- F(m^{2}_{H^{0}}, m^{2}_{A^{0}})\right],\\
\Delta U_{ID} &= 0.
\end{align}
where, the $F(x,y)$ function is given by
\begin{equation}
F(x, y) =
    \begin{cases}
    \frac{x+y}{2} - \frac{xy}{x-y} \log \left( \frac{x}{y} \right), & \text{for  $x \neq y$},\\
    0,  & \text{for $x = y$}.
    \end{cases}
\end{equation}

For the inert triplet~\cite{Forshaw_2001}
\begin{align}
\Delta S_{IT} &= 0, \\
\Delta T_{IT} &= \dfrac{1}{8\pi} \dfrac{1}{s^{2}_{W}M^{2}_{W}}\left[m^{2}_{T^{+}} + m^{2}_{T^{0}}- \dfrac{2m^{2}_{T^{+}}m^{2}_{T^{0}}}{m^{2}_{T^{+}} - m^{2}_{T^{0}}}\log \left(\dfrac{m^{2}_{T^{+}}}{m^{2}_{T^{0}}} \right) \right], \\
\begin{split}
\Delta U_{IT} &= -\dfrac{1}{3\pi}\left[m^{2}_{T^{+}}\log \left(\dfrac{m^{2}_{T^{+}}}{m^{2}_{T^0}} \right)\dfrac{3m^{2}_{T^{0}}-m^{2}_{T^{+}}}{(m^{2}_{T^{+}}-m^{2}_{T^{0}})^{3}}\right]\\
&-\dfrac{1}{3\pi} \left[\dfrac{5(m^{4}_{T^{+}} + m^{4}_{T^{0}})-22m^{2}_{T^{+}}m^{2}_{T^{0}}}{6(m^{2}_{T^{+}}-m^{2}_{T^{0}})^{2}}   \right].
\end{split}
\end{align}

Finally, for the active triplet, we obtain~\cite{CHENG2023116118}
\begin{align}
\begin{split}
\Delta S_{AT} & =  -\dfrac{1}{3\pi}\log \left( \dfrac{m^{2}_{\Delta^{++}}}{m^{2}_{\Delta^{0}}}\right)  - \dfrac{2}{\pi}\left[(1-2s^{2}_{W})^{2}\xi\left(\dfrac{m^{2}_{\Delta^{++}}}{m^{2}_{Z}}, \dfrac{m^{2}_{\Delta^{++}}}{m^{2}_{Z}}\right)\right]\\
&- \dfrac{2}{\pi}\left[s^{4}_{W}\xi \left( \dfrac{m^{2}_{\Delta^{+}}}{m^{2}_{Z}}, \dfrac{m^{2}_{\Delta^{+}}}{m^{2}_{Z}} \right) 
 +  \xi\left( \dfrac{m^{2}_{\Delta^{0}}}{m^{2}_{Z}}, \dfrac{m^{2}_{\Delta^{0}}}{m^{2}_{Z}}\right) \right],
\end{split}
\\
\Delta T_{AT} &= \dfrac{1}{4\pi s^{2}_{W}M^{2}_{W}}\left[F(m^{2}_{\Delta^{++}}, m^{2}_{\Delta^{+}})+ F(m^{2}_{\Delta^{+}}, m^{2}_{\Delta^{0}})\right],\\
\begin{split}
\Delta U_{AT} &=  \dfrac{1}{6\pi}\log \left(\dfrac{m^{4}_{\Delta^{+}}}{m^{2}_{\Delta^{++}}m^{2}_{\Delta^{0}}}\right) \\
& + \dfrac{2}{\pi}\left[(1-2s^{2}_{W})^{2}\xi\left(\dfrac{m^{2}_{\Delta^{++}}}{m^{2}_{Z}}, \dfrac{m^{2}_{\Delta^{++}}}{m^{2}_{Z}}\right) + s^{4}_{W}\xi \left( \dfrac{m^{2}_{\Delta^{+}}}{m^{2}_{Z}}, \dfrac{m^{2}_{\Delta^{+}}}{m^{2}_{Z}} \right)
+  \xi\left( \dfrac{m^{2}_{\Delta^{0}}}{m^{2}_{Z}}, \dfrac{m^{2}_{\Delta^{0}}}{m^{2}_{Z}}\right) \right] \\
&- \dfrac{2}{\pi}\left[ \xi\left( \dfrac{m^{2}_{\Delta^{++}}}{m^{2}_{W}}, \dfrac{m^{2}_{\Delta^{+}}}{m^{2}_{W}} \right) + \xi \left(\dfrac{m^{2}_{\Delta^{+}}}{m^{2}_{Z}}, \dfrac{m^{2}_{\Delta^{0}}}{m^{2}_{Z}} \right) \right].
\end{split}
\end{align}

Where the functions $\xi(x,y)$, $d(x,y)$ and $f(x,y)$ are defined as follows
\begin{align*}
    \xi(x,y) &= \dfrac{4}{9} - \dfrac{5}{12}(x+y) + \dfrac{1}{6}(x-y)^{2} + \dfrac{1}{4}\left[x^{2}-y^{2} - \dfrac{1}{3}(x-y)^{3}-\dfrac{x^{2}+y^{2}}{x-y}\right]\log\left(\dfrac{x}{y}\right) \\
    &- \dfrac{1}{12}\,d(x,y)\,f(x,y) \\
    d(x,y) &= -1 + 2(x+y) - (x-y)^{2}\\
    f(x,y) &=
    \begin{cases}
    -2\sqrt{d(x,y)}\left[\arctan\dfrac{x- y + 1}{\sqrt{d(x,y)}} - \arctan\dfrac{x - y - 1}{\sqrt{d(x,y)}}\right], & \text{  $d(x,y) > 0$} \\
    \sqrt{-d(x,y)}\log\left( \dfrac{x+ y-1 +\sqrt{-d(x,y)} }{x + y -1 - \sqrt{-d(x,y)}}\right), & \text{  $d(x,y) \leq 0$} 
    \end{cases}.
\end{align*}

We have performed a numerical scan of the contribution to the oblique parameters $S$, $T$ and $U$ of every field in our model. Considering the experimental constraints on the masses of the scalars, we have used random values within the ranges: $100 \leq m_{H^{0}}\leq 1000$~GeV for the complex inert doublet, $100\leq m_{T^{0}}\leq 2500$~GeV for the real inert triplet and $850 \leq m_{\Delta^0}\leq 3500$~GeV for the real active triplet. Furthermore, the mass splitting between the charged and neutral components are set as: $m_{H^{\pm}}-m_{H^{0}}= m_{A^{0}}-m_{H^{0}} = 1$~GeV for the inert doublet, $m_{\Delta^{\pm\pm}}-m_{\Delta^{\pm}} = m_{\Delta^{\pm}}- m_{\Delta^{0}} = 10$~GeV for the active triplet and $\Delta m = 166$~MeV(eq.~\ref{eq:tripletsplitting}) for the inert triplet. It is worth to mention that these values are in agreement with the experimental constraints in Section~\ref{Sec:Constraints}.

\begin{figure}[htb]
    \centering
     \begin{subfigure}[b]{0.47\textwidth}
      \includegraphics[scale=0.45]{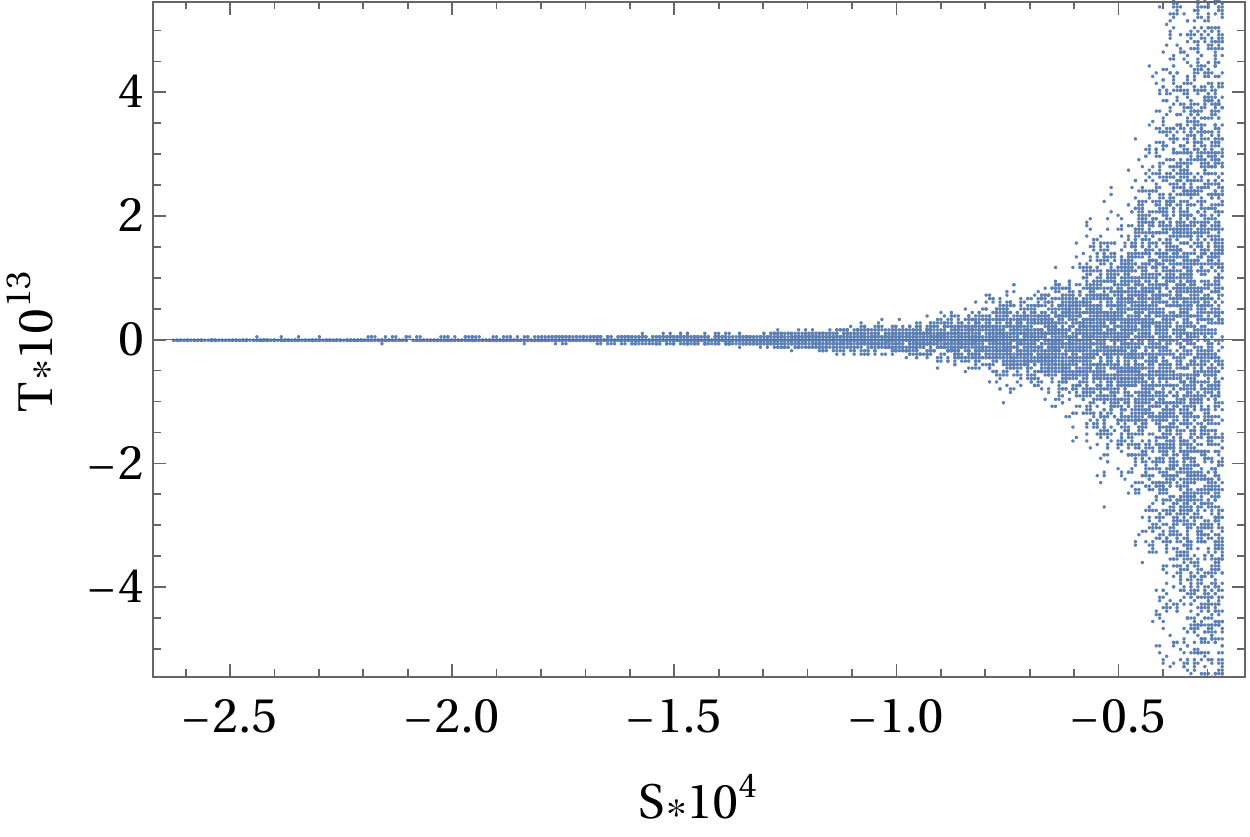}
      \caption{}
     \end{subfigure}
     \begin{subfigure}[b]{0.47\textwidth}
      \includegraphics[scale=0.485]{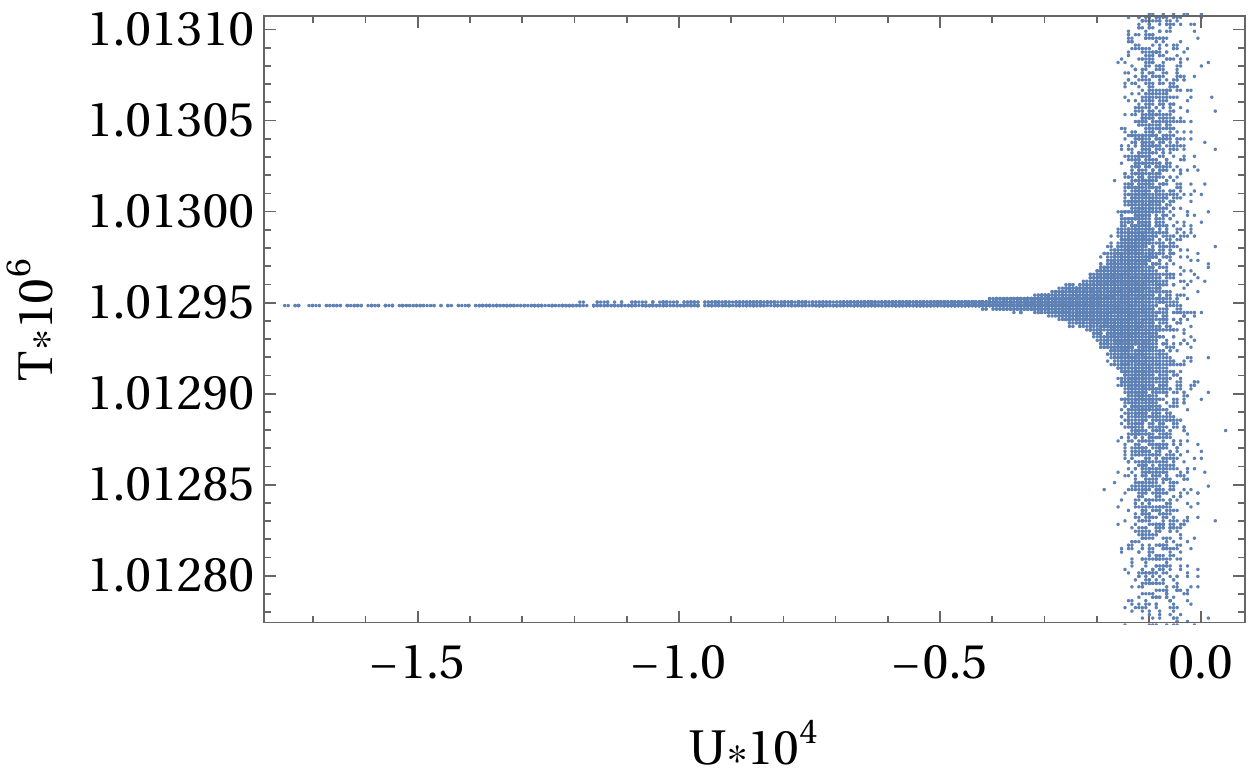}
      \caption{}
     \end{subfigure}
     \begin{subfigure}[b]{0.47\textwidth}
      \includegraphics[scale=0.48]{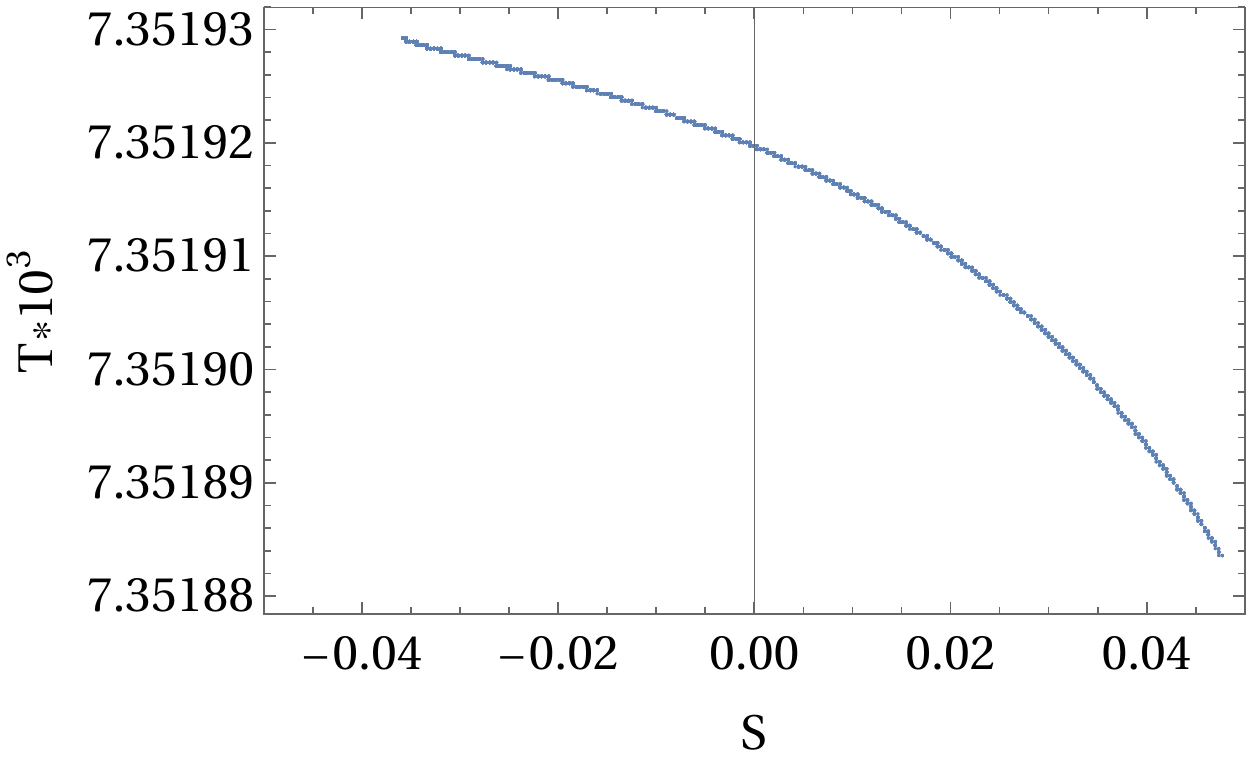}
      \caption{}
     \end{subfigure}
     \begin{subfigure}[b]{0.47\textwidth}
      \includegraphics[scale=0.48]{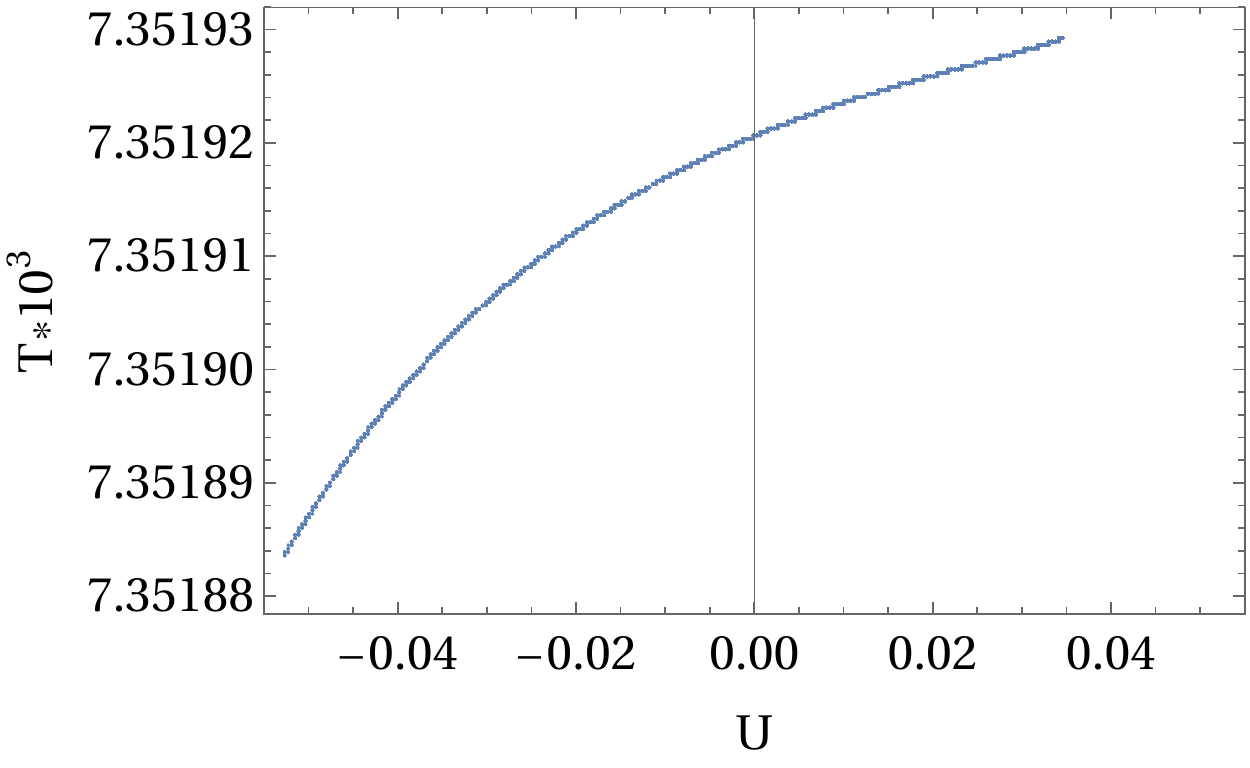}
      \caption{}
     \end{subfigure}
    \caption{Contribution to the oblique parameters $S, T$ and $U$ for (a) Inert doublet, (b) Inert triplet, (c) Active triplet and (d) Active triplet}
    \label{fig:obliqueparameters}
\end{figure}
We show in Figure~\ref{fig:obliqueparameters} the results for the $S$, $T$ and $U$ oblique parameters for each field, which lie within the most recent bounds~\cite{PDG2022}:
\begin{align}
    \Delta S & = - 0.02 \pm 0.10, \\
    \Delta T & =  0.03 \pm 0.12, \\
    \Delta U & =  0.01 \pm 0.11.
\end{align}
It is worth to notice that the biggest contribution comes from the real active triplet, as expected, while the contribution of the inert fields are practically negligible in comparison. Therefore, the whole mass spectra range for the scalar fields considered is safely allowed by the bounds to the total $S$, $T$ and $U$ oblique parameters.

\section{Results}\label{Sec:Results}

\begin{figure}[htb]
    \centering
     \begin{subfigure}[b]{0.47\textwidth}
      \includegraphics[scale=0.46]{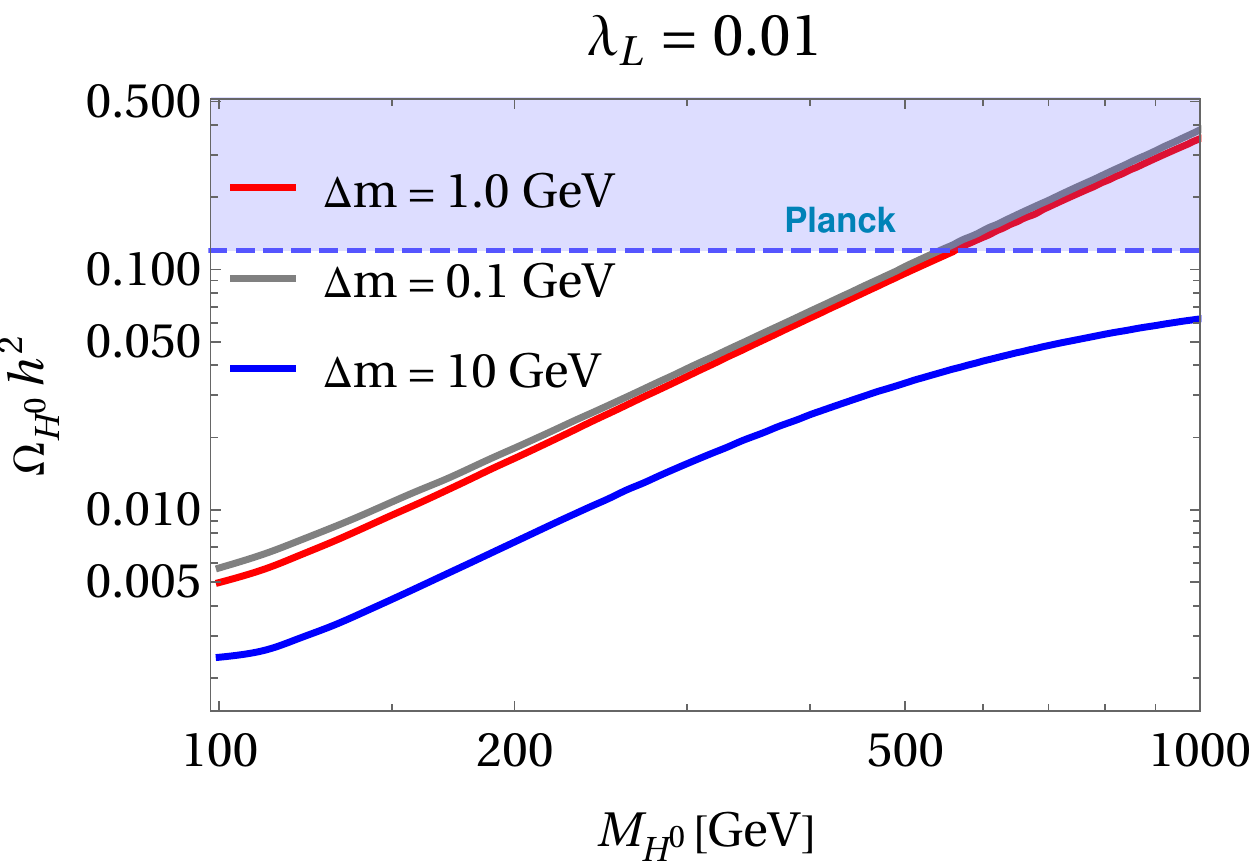}
      \caption{}
      \label{IDM}
     \end{subfigure}
     \begin{subfigure}[b]{0.47\textwidth}
      \includegraphics[scale=0.47]{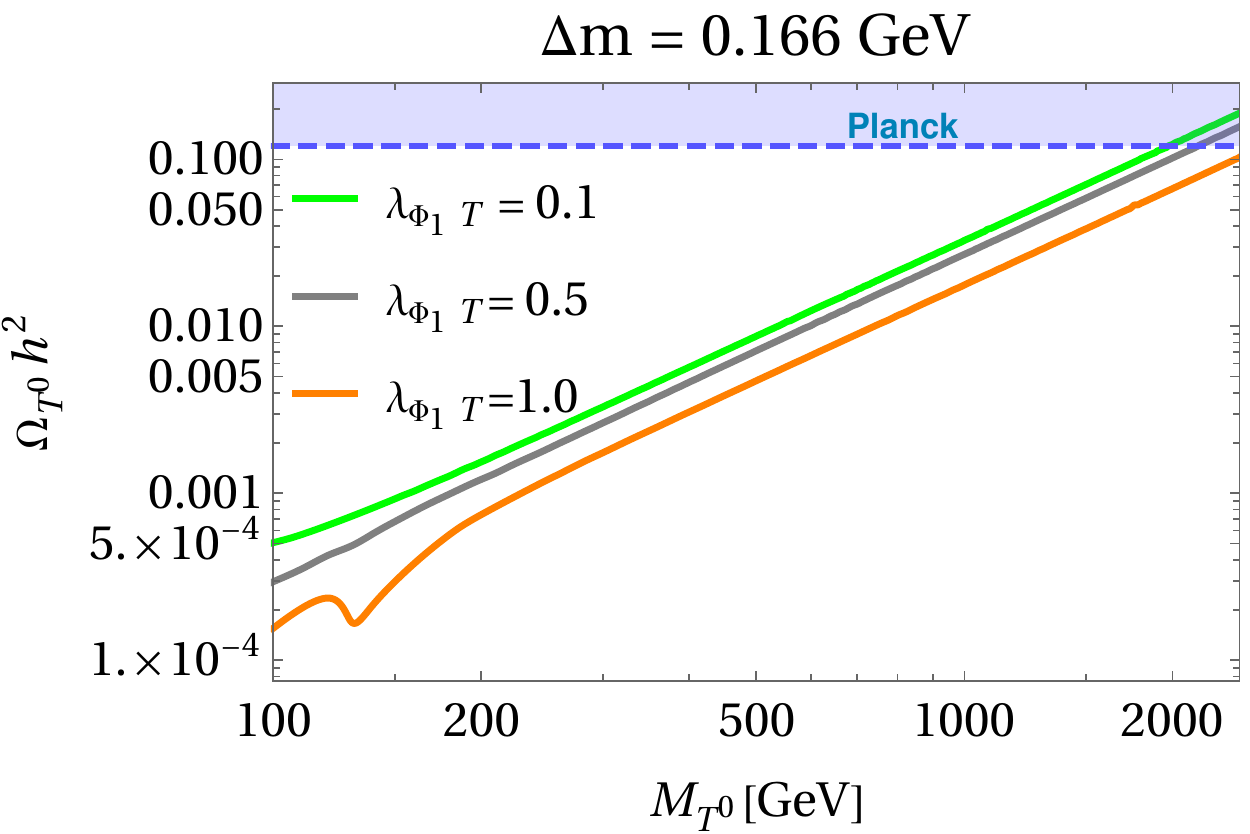}
      \caption{}
      \label{ITM}
     \end{subfigure}
    \caption{Relic density in the IDM (left) and ITM (right). The dashed line shows the Planck limit~\cite{PLANCK2018} for the DM abundance.}
    \label{fig:IDMandITM}
\end{figure}

As we mention in section~\ref{sec:Model}, the relevant parameters for the DM phenomenology in the model are: $m_{H^{0}}$, $m_{H^{\pm}}$, $m_{A^{0}}$, $m_{T^{0}}$, $m_{T^{\pm}}$, $\lambda_{L}$, $\lambda_{\Phi_{1}T}$ and $\lambda_{\Phi_{2}T}$. The other parameters do not take a role in the relic density calculation; consequently, we have fixed them in the following manner: $\lambda_{\Phi_{1}\Delta} = \lambda_{\Phi_{2}\Delta}$, $\lambda_{\Delta} = 0.01$, $\lambda_{T} = 0.1$, $m_{\Delta^{0}} = 1000$ GeV, $m_{\Delta^{\pm\pm}}- m_{\Delta^{\pm}} = m_{\Delta^{\pm}}-m_{\Delta^{0}} = 10$ GeV and $\kappa_{\Phi_{1}\Delta} = \kappa_{\Phi_{2}\Delta} \simeq 10^{-8}$ GeV. To perform our analysis we have used the following computer tools, SARAH~\cite{STAUB20141773} to implement the model and MicrOMEGAs~\cite{MicrOmegas2DMC} to compute the dark matter relic abundance and Spin-Independent DM nucleon cross-section for each candidate.

\subsection{Relic Density}

First, we have computed the relic density for the inert doublet and the inert triplet for each individual model in figure~\ref{fig:IDMandITM}. The left panel displays the relic density for the ID as a function of the DM candidate mass $H_{0}$. The important parameters for relic density calculation in this model are $m_{H^{0}}$, $\lambda_{L}$ and the mass splitting $\Delta m$ between the scalars $A_{0}$ and $H_{0}$.  In this case, we have plotted the relic density for different values of $\Delta m$ while keeping $\lambda_{L} = 0.01$. We observe that the desert region extends to $550$ GeV when $\Delta m$ is small $\leq 1$ GeV. Further, we notice that changing $\Delta m$ between $0.1$ GeV and $1.0$ GeV does not lead to significantly different results, while larger values of $\Delta m$ suppress the relic density due to contribution of electroweak gauge bosons annihilation~\cite{SARMA2021115300}. For instance, one notices that setting $\Delta m = 10$ GeV turns the entire mass range under-abundant. It is worth mentioning that an appropriate choice of $\lambda_{L}$ is needed to obtain the correct relic density. $\lambda_{L}$ must be small to avoid under-abundance. In the following, we have fixed $\lambda_{L} = 0.01$ and $\Delta m = 1$ GeV.

On the other hand, the desert region for the IT extends up to $1970$~GeV. This occurs because of the very small mass splitting (see eq.~\ref{eq:tripletsplitting}) between the charged $T^{\pm}$ and neutral scalars $T^{0}$, which results in a larger annihilation cross section. In addition, we observe that the relic density for the inert triplet is larger when the Higgs portal coupling $\lambda_{\Phi_{1}T} = 0.1$. For $\lambda_{\Phi_{1}T}= 1.0$, the relic density is significantly smaller. Therefore, in order to maximise the relic density for $T_{0}$, we have fixed $\lambda_{\Phi_{1}T} = 0.1$. 

\begin{figure}[htb]
    \centering
    \begin{subfigure}{0.4\textwidth}
    \centering
     \includegraphics[scale=0.25]{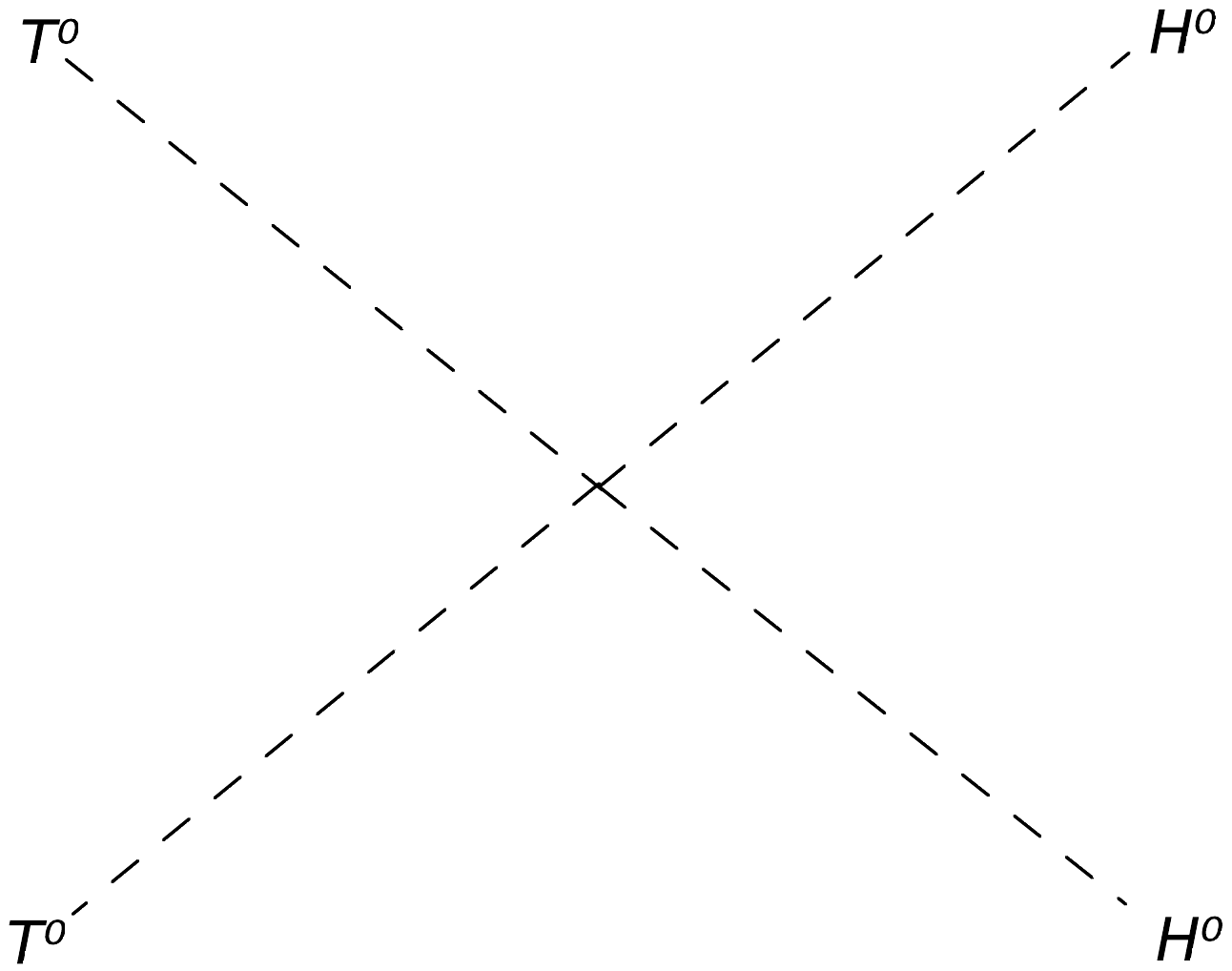}
     \caption{}
    \end{subfigure}
     \begin{subfigure}{0.4\textwidth}
     \centering
     \includegraphics[scale=0.25]{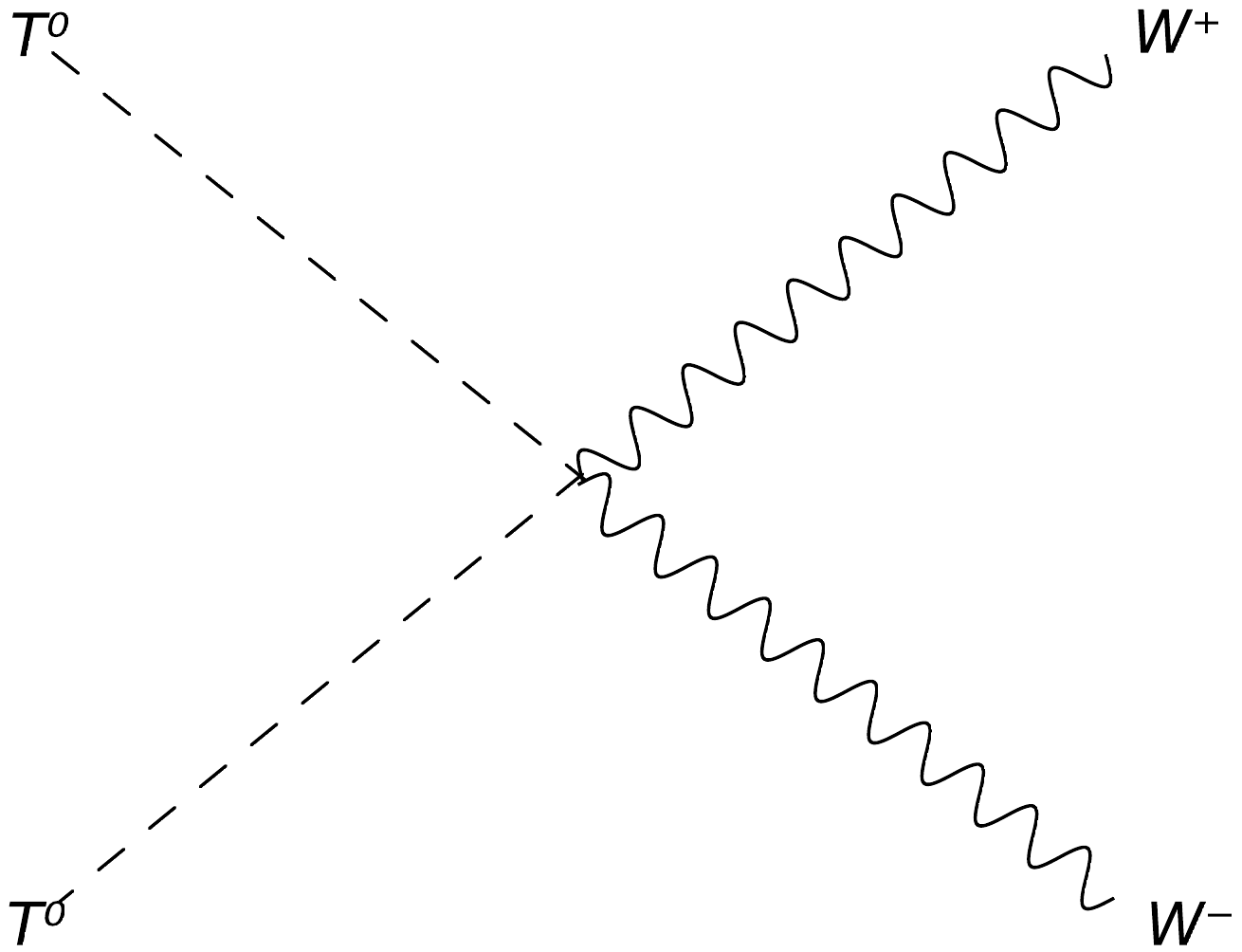}
     \caption{}
    \end{subfigure}
     \begin{subfigure}{0.4\textwidth}
     \centering
     \includegraphics[scale=0.25]{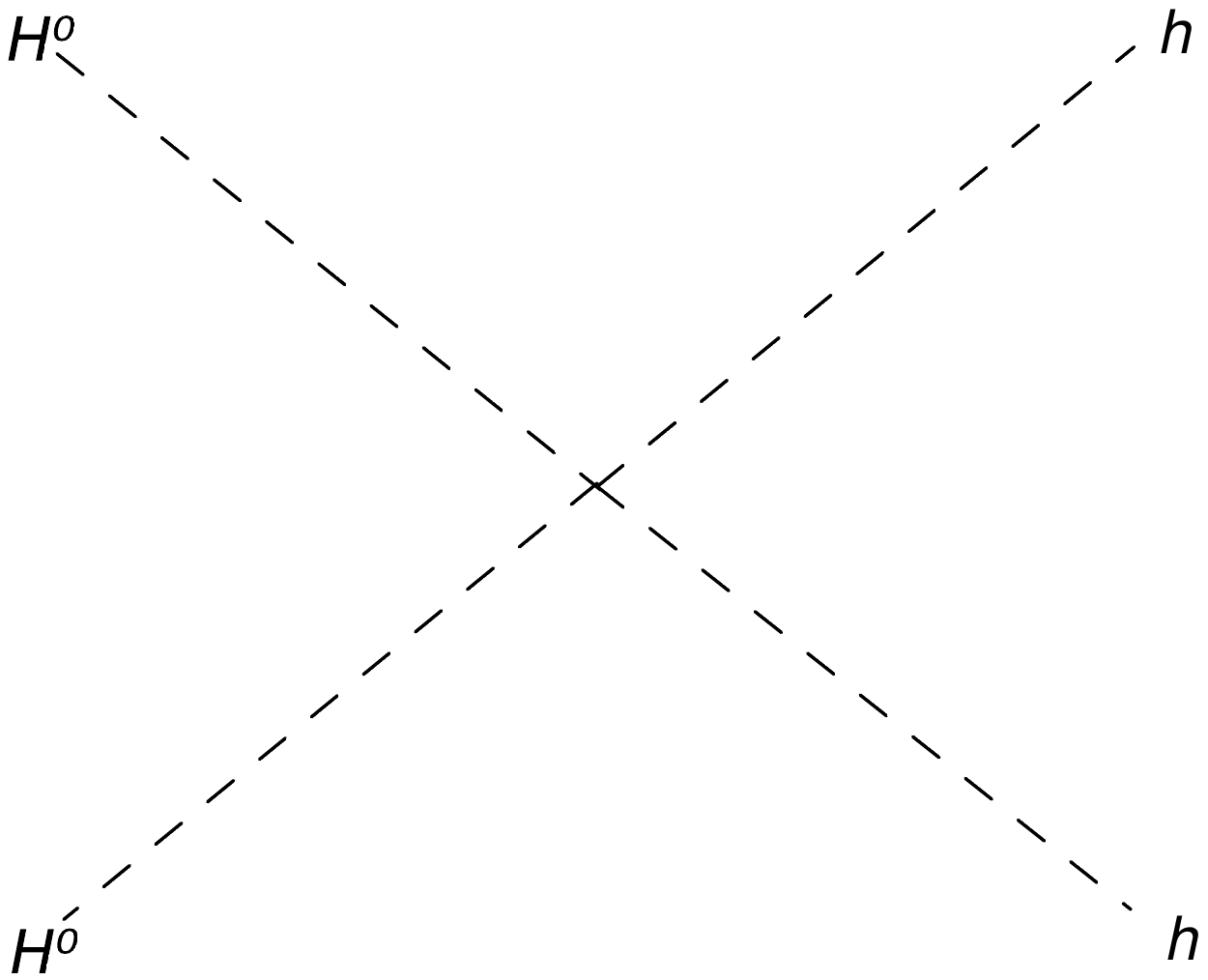}
     \caption{}
    \end{subfigure}
     \begin{subfigure}{0.4\textwidth}
     \centering
     \includegraphics[scale=0.25]{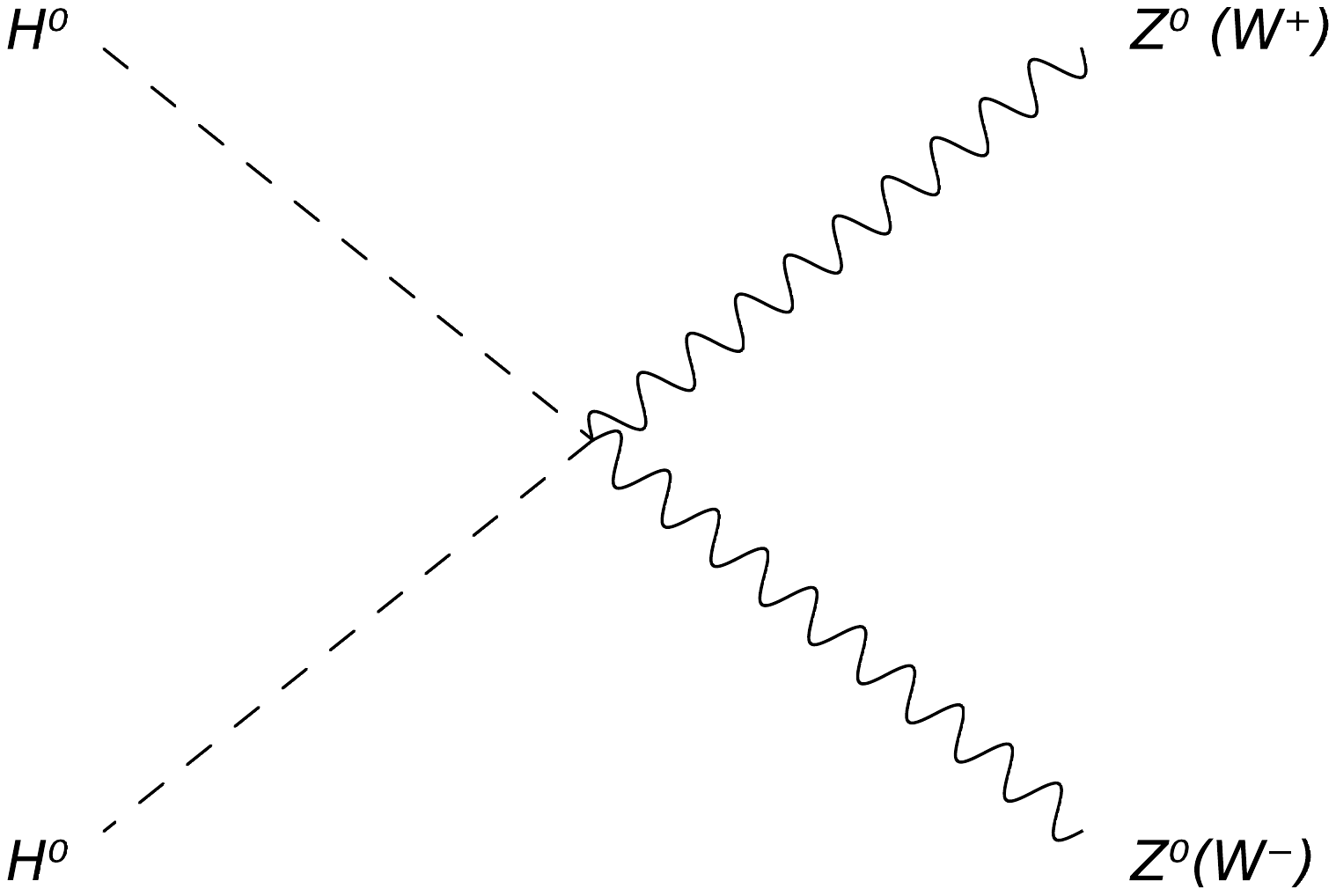}
     \caption{}
    \end{subfigure}
    \begin{subfigure}{0.4\textwidth}
    \centering
     \includegraphics[scale=0.25]{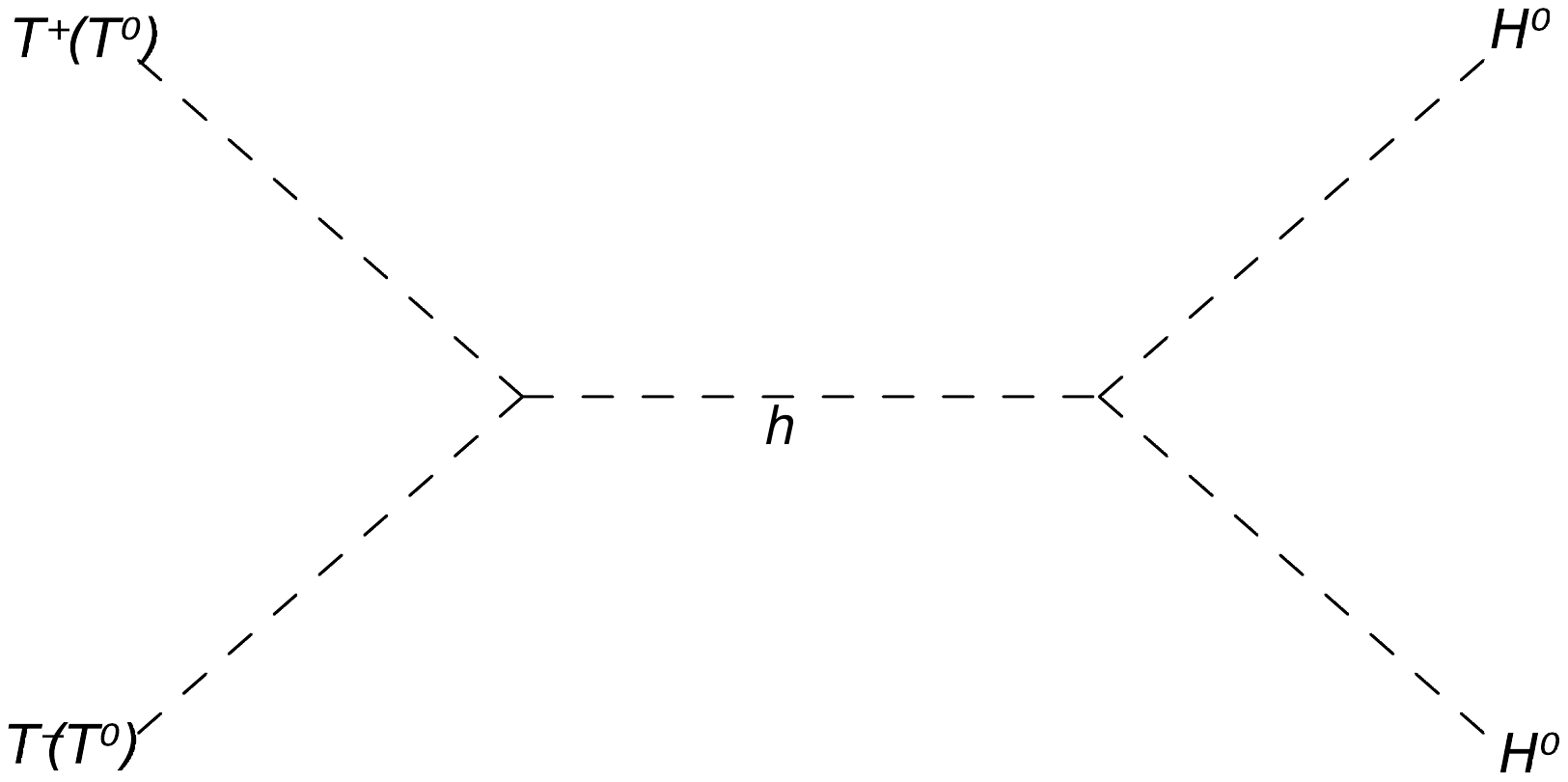}
     \caption{}
    \end{subfigure}
    \begin{subfigure}{0.4\textwidth}
    \centering
     \includegraphics[scale=0.25]{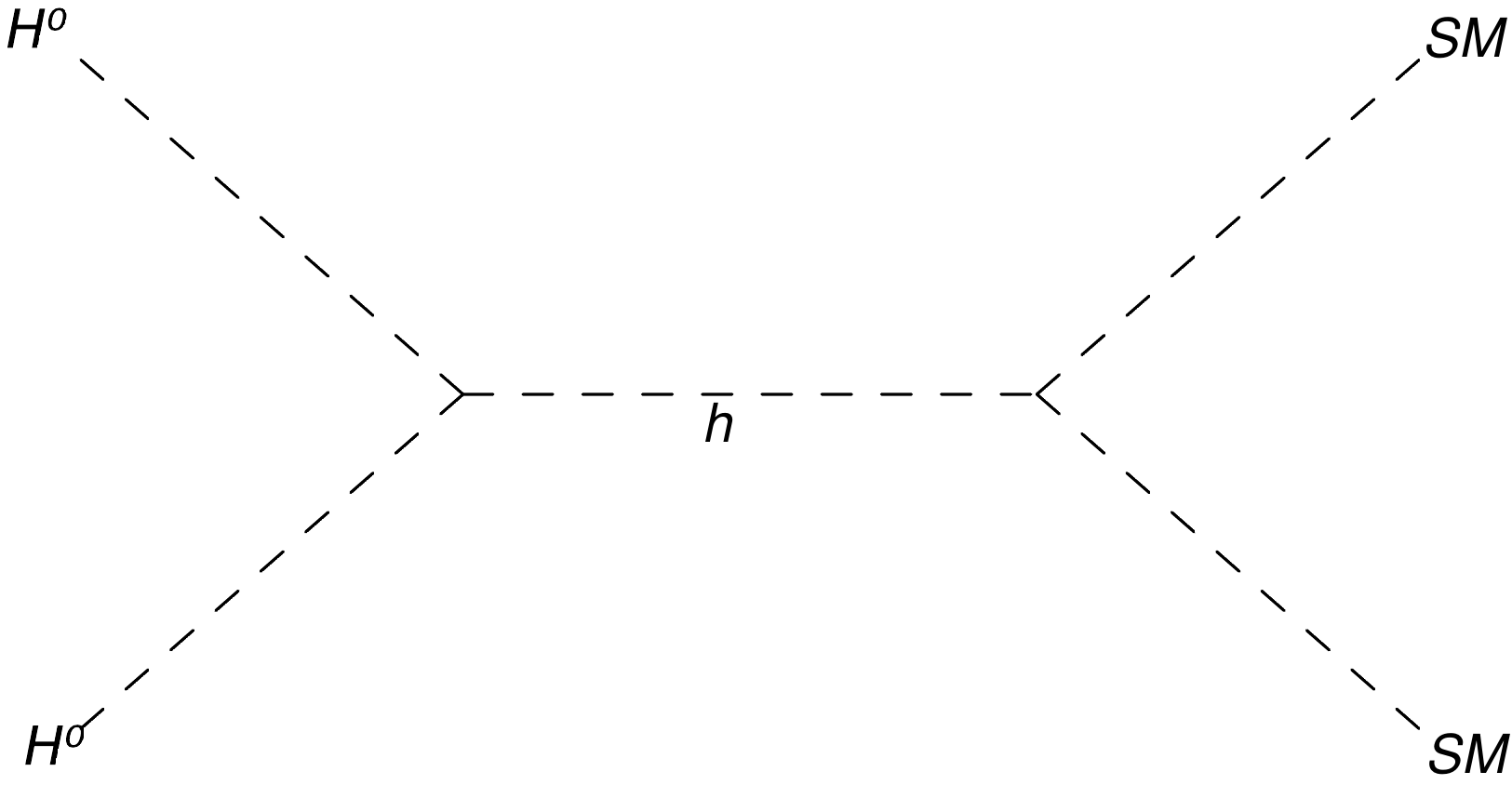}
     \caption{}
    \end{subfigure}
    \caption{Main annihilation channels contributing to the DM annihilation cross-section for $m_{T^0} > m_{H^0}$: (a) DM-DM conversion, (b)  DM-W bosons, (c) DM-Higgses, (d) DM-$W^{\pm}$ bosons ($Z^{0}$ bosons), (e) DM-DM conversion from s-channel and (f) DM annihilation into SM particles.}
    \label{fig:DMdiagrams}
\end{figure}

We aim to reach the correct total relic density through the contribution of a second DM candidate, the IT, which belongs to a different dark sector.
In our analysis we have considered $m_{T^0} > m_{H^0}$; the main annihilation channels that contribute to the annihilation cross-section are shown in figure~\ref{fig:DMdiagrams}. The region of interest for each candidate is: 100~GeV $\leq m_{H^{0}} \leq $ 550~GeV for the ID and 100~GeV $\leq m_{T^{0}} \leq $ 2500~GeV for the IT. It is worth pointing out that these regions lie within the allowed mass limits in colliders (see eq.~\ref{eq:constraintsneu} and eq.~\ref{eq:constraintschar}).

\begin{figure}[htb]
    \centering
     \begin{subfigure}[b]{0.46\textwidth}
      \includegraphics[scale=0.46]{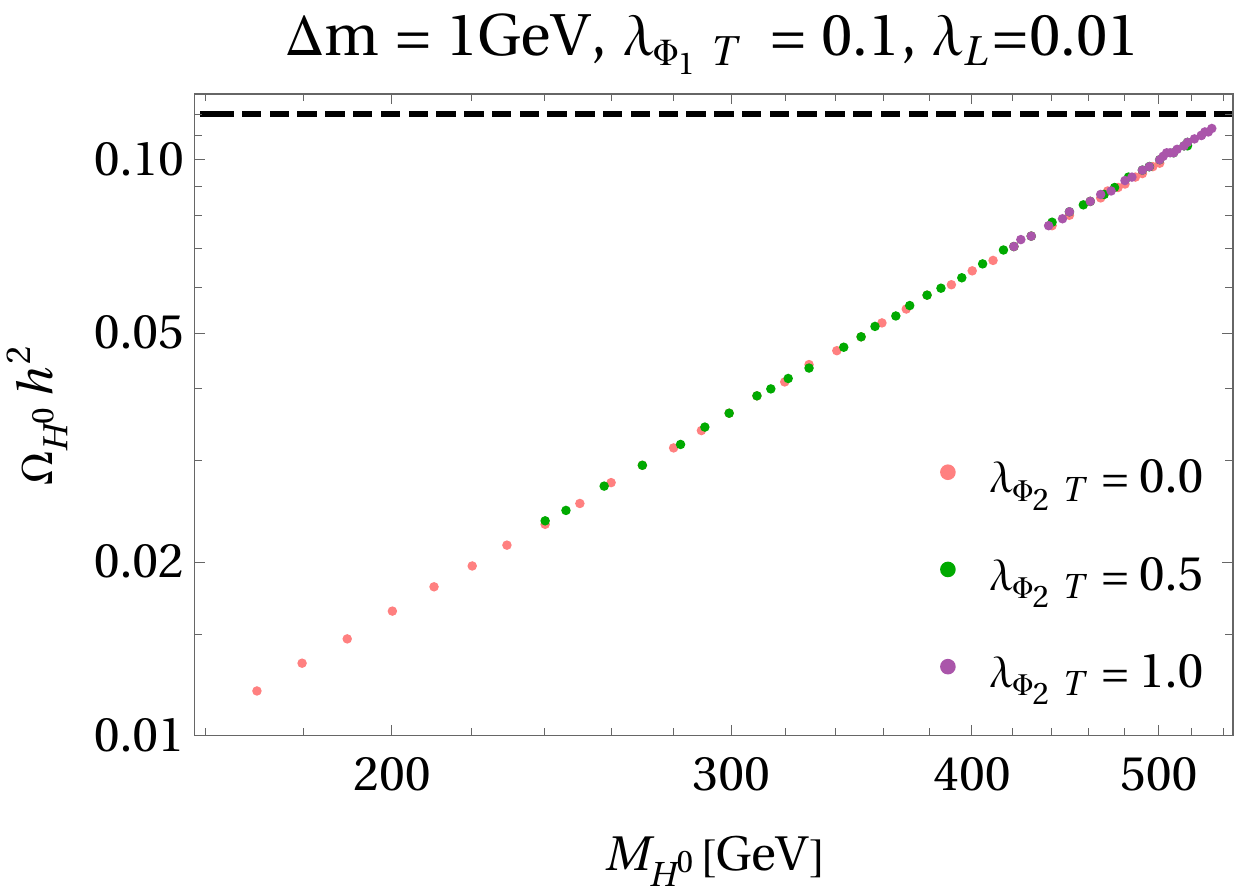}
      \caption{}
     \end{subfigure}
     \begin{subfigure}[b]{0.46\textwidth}
      \includegraphics[scale=0.47]{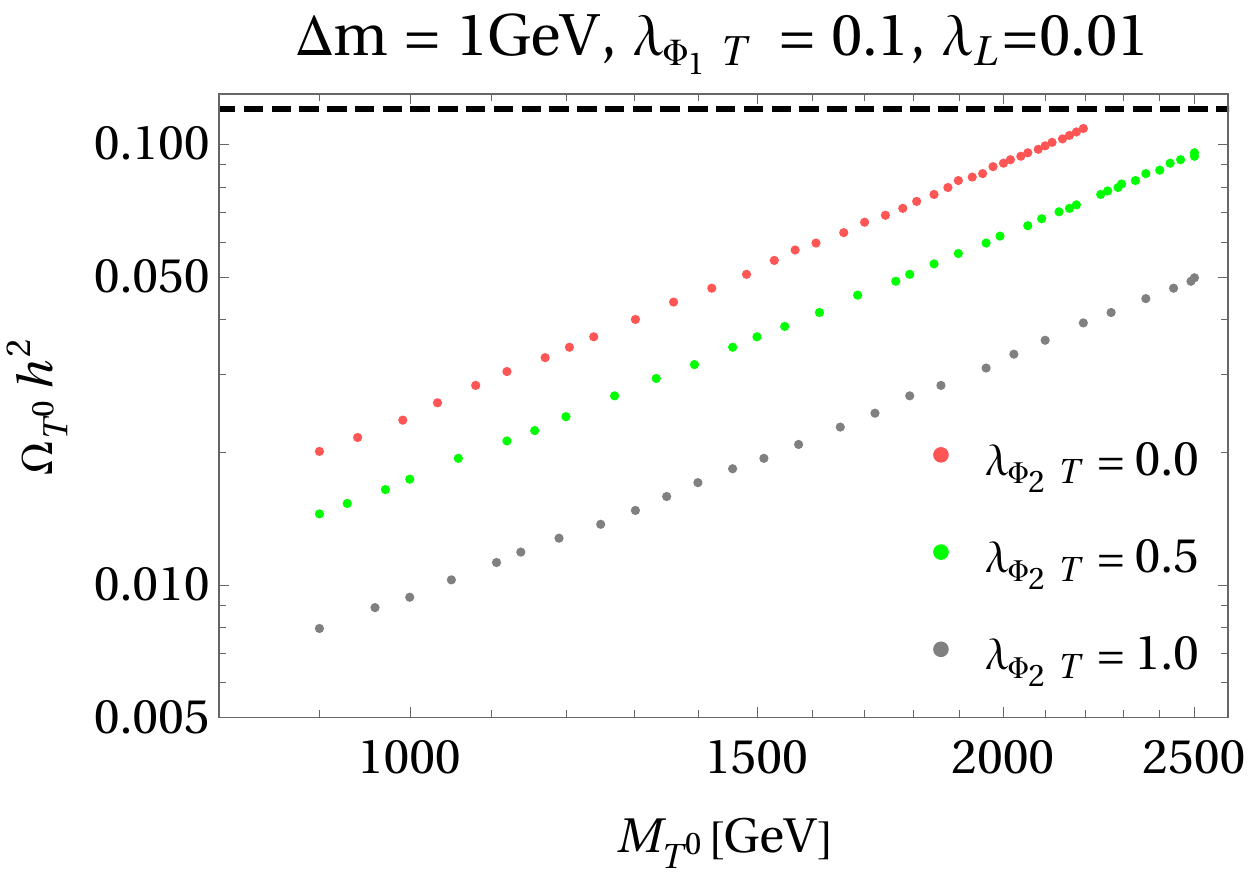}
      \caption{}
     \end{subfigure}
    \caption{Contribution of each candidate to the total relic density $\Omega_{H^{0}} h^{2} + \Omega_{T^{0}} h^{2} = \Omega_{T} h^{2}$ for $\lambda_{\Phi_{2}T} = \{0.0, 0.5, 1.0\}$, $\lambda_{L} = 0.01$, $\Delta m = 1$ GeV and $\lambda_{\Phi_{1}T} = 0.1$. The points satisfy the Planck limit~\cite{PLANCK2018} for the relic density of DM (dashed line).}
    \label{fig:RelicDensityVsDMMass}
\end{figure}

In order to study the effect of the conversion from $T^{0}$ into $H^{0}$, we have computed the relic density for each DM candidate as a function of the mass for different values of $\lambda_{\Phi_2 T} = \{0, 0.5, 1.0\}$. In figure~\ref{fig:RelicDensityVsDMMass}, we have shown our results for the relic density of each candidate, while maintaining $M_{T^0} > M_{H^0}$. The dashed line in each plot represents the Planck limit for the DM over-abundance. We observe that the relic density for each candidate is again under-abundant for most of the mass range. However, the contribution of both candidates $\Omega_{Tot} h^{2} = \Omega_{H^{0}}h^{2} + \Omega_{T^{0}}h^{2}$ satisfies the Planck limit for each value of the $\lambda_{\Phi_{2}T}$ coupling.

\subsection{Direct Detection}

\begin{figure}[htb]
    \centering
      \begin{subfigure}[b]{0.47\textwidth}
      \centering
       \includegraphics[scale=0.5]{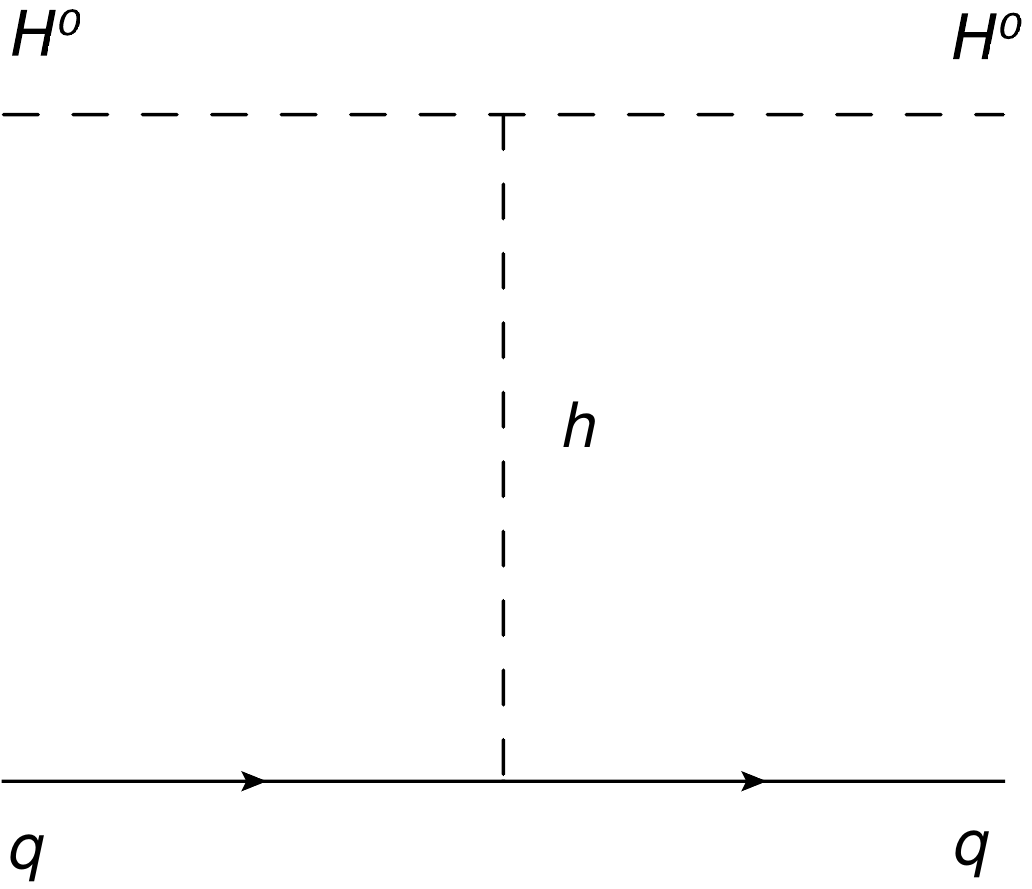}
       \caption{}
       \label{diagram1}
      \end{subfigure}
      \begin{subfigure}[b]{0.47\textwidth}
      \centering
       \includegraphics[scale=0.5]{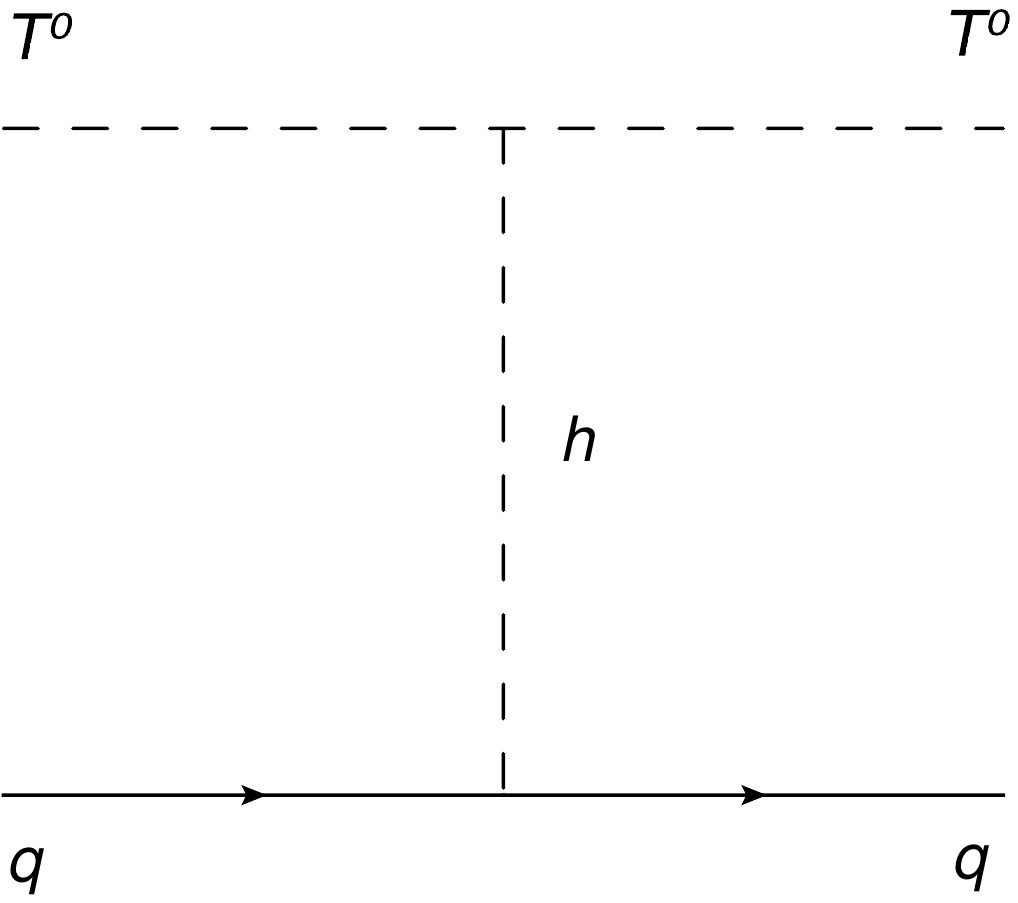}
       \caption{}
       \label{diagram2}
      \end{subfigure}
    \caption{Tree-level Spin-Independent DM-nucleon elastic scattering channels for each DM candidate.}
    \label{dispersion}
\end{figure}

In addition to the relic density constraint, we have considered the experimental upper limits for the Spin-Independent (SI) cross-section of DM-nucleons interactions. The corresponding Feynman diagrams for the elastic scattering of DM and nucleons are shown in figure~\ref{dispersion}. 
We have computed the SI cross-section for each candidate as a function of its mass using micrOMEGAs. In order to compare our results with the experimental limits, we take in account that those limits are based on one-component DM scenarios. Therefore, we need to rescale the cross section for each component in our model as: $\sigma_{SI}^{i}*\Omega_{i} h^{2}/\Omega_{\text{Tot}} h^{2}$, where $i = H_{0},~T_{0}$.

\begin{figure}[htb]
    \centering
      \begin{subfigure}[b]{0.47\textwidth}
       \includegraphics[scale=0.47]{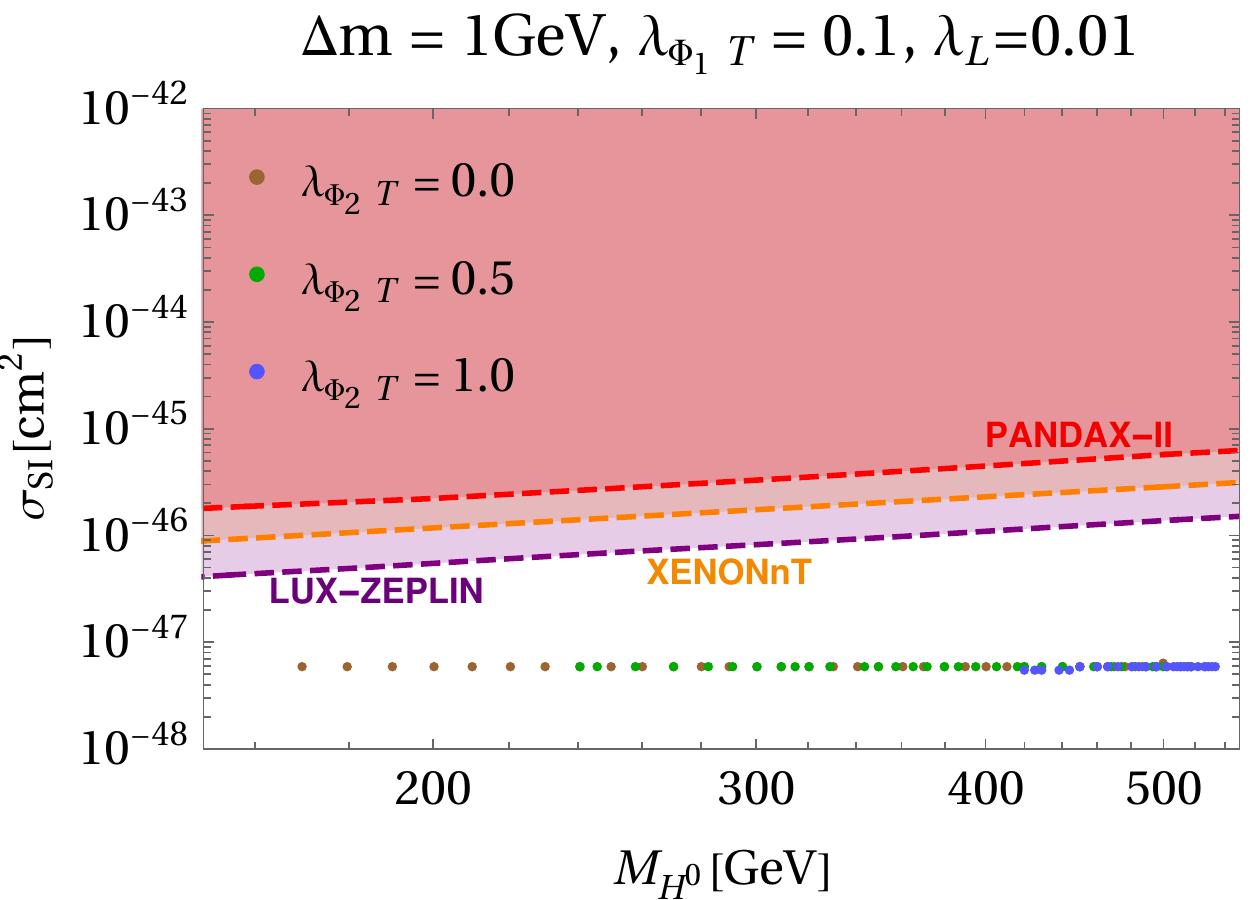}
       \caption{}
      \end{subfigure}
      \begin{subfigure}[b]{0.47\textwidth}
       \includegraphics[scale=0.48]{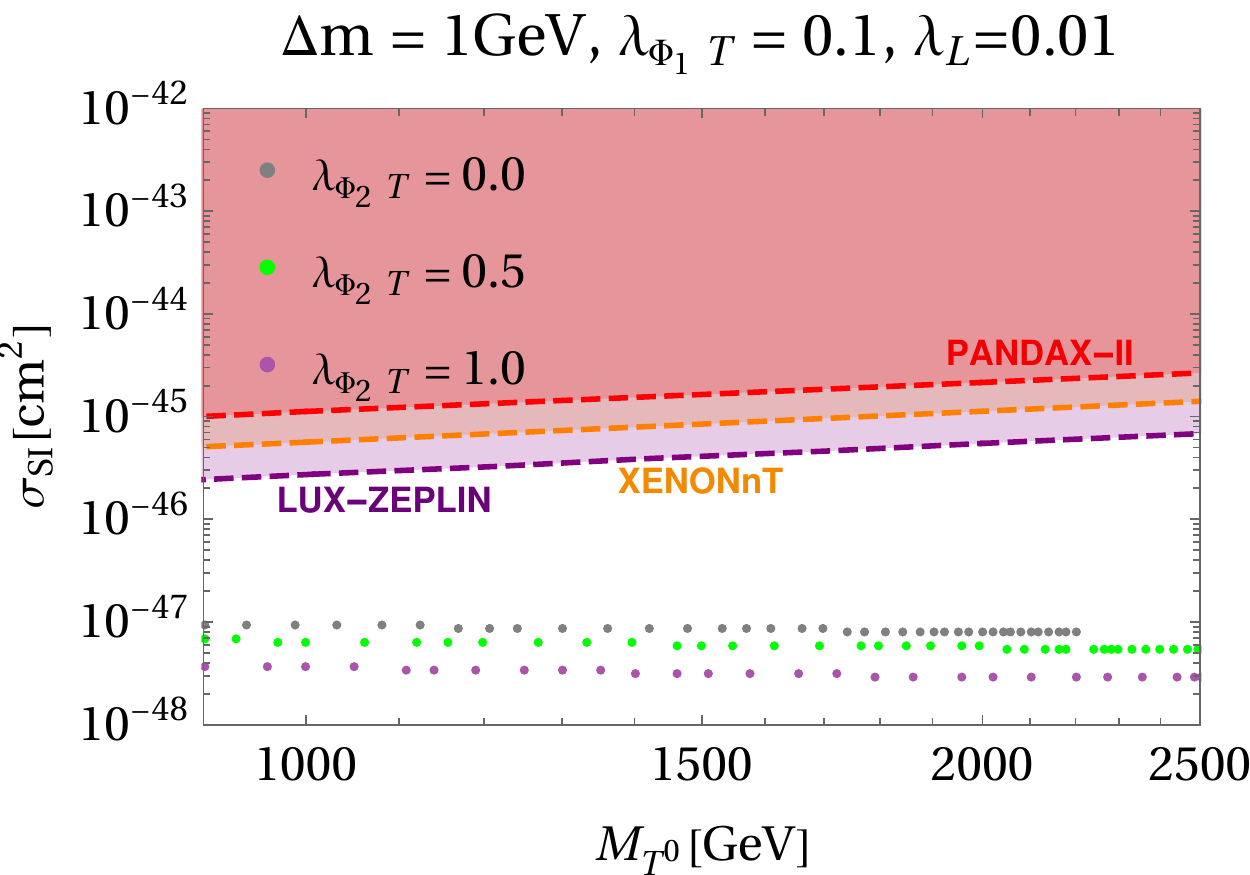}
       \caption{}
      \end{subfigure}
   \caption{Spin-Independent cross-section of DM-nucleon interaction for (a) $H^0$ and (b) $T^0$ for $\lambda_{\Phi_{2}T} = \{0.0, 0.5, 1.0\}$, $\lambda_{L} = 0.01$, $\Delta m = 1$ GeV and $\lambda_{\Phi_{1}T} = 0.1$. Dashed lines represent the current limits of PANDAX-II~\cite{PANDA2017}, XENONnT~\cite{XENONnT2023} and LUX-ZEPLIN~\cite{LZ2023} experiments. Shaded areas are the regions excluded by those experiments. Points represent our results.}
    \label{crosssectionDD}
\end{figure}

Figure~\ref{crosssectionDD} shows the SI cross-section for the DM-nucleon interaction as a function of the mass of each DM candidate. The different dashed lines indicate the upper experimental limits of the PANDAX-II, XENONnT and LUX-ZEPLIN collaborations.
The pink and orange shaded regions are excluded by these limits. The solid lines represent our results for different values of $\lambda_{\Phi_{2}T} = \lbrace 0.0, 0.5, 1.0 \rbrace$. All points satisfy the theoretical constraints as well as the correct relic density. We observe that our results for both candidates are within the experimental limits. Furthermore, it is worth mentioning that our results lie above the expected neutrino-nucleus coherent scattering limit for direct detection dark matter experiments~\cite{Essig:2018tss}.

\begin{figure}[htb]
\centering
\includegraphics[scale=0.64]{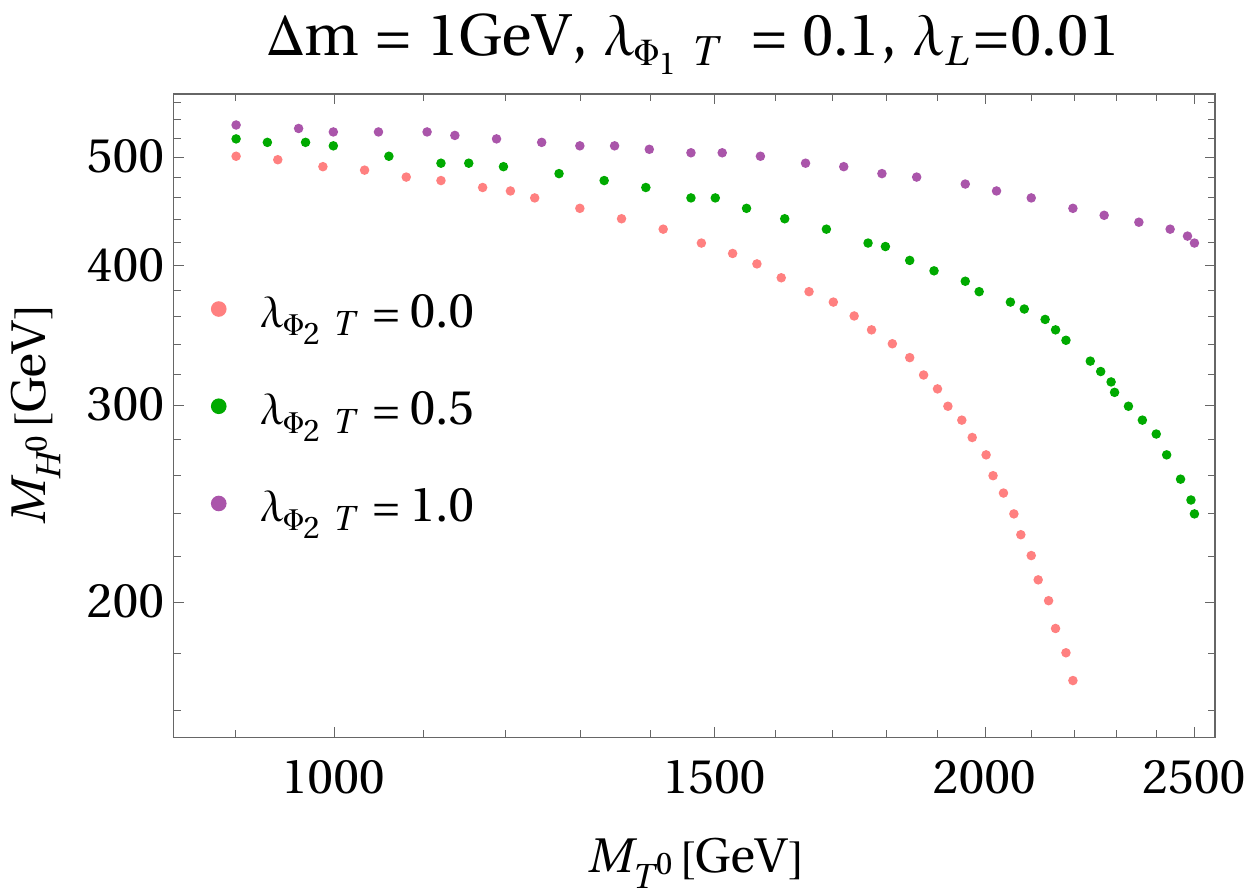}
\caption{Points allowed by all theoretical constraints that also satisfy the correct relic density and direct detection limits for $\lambda_{\Phi_{2}T} = \{0, 0.5, 1.0\}$.}
\label{relicpoints}
\end{figure}

In table~\ref{tb:3} we have shown some benchmarks allowed by all theoretical and experimental constraints considered. We observe that for the selected values of $\lambda_{\Phi_{2}T}$, the total relic density $\Omega_{Tot}h^{2}$ lies within the Planck limit. All the points allowed by all theoretical and experimental constraints including relic density and direct detection limits are shown in figure~\ref{relicpoints} in the $M_{T^{0}}-M_{H^{0}}$ plane. Considering the desert zone of the inert doublet, $100\leq m_{H^{0}} < 550$ GeV, we observe that almost the entire mass range is now allowed $170 \leq m_{H^{0}} \leq 550$ GeV. For the inert triplet the correct total relic density is satisfied within the region $900 \leq m_{T^{0}} \leq 1970$ GeV, making available a significant portion of the desert region ($100\leq m_{T^{0}} < 1970$~GeV).

\begin{figure}[htb]
    \centering
      \begin{subfigure}[b]{0.47\textwidth}
       \includegraphics[scale=0.48]{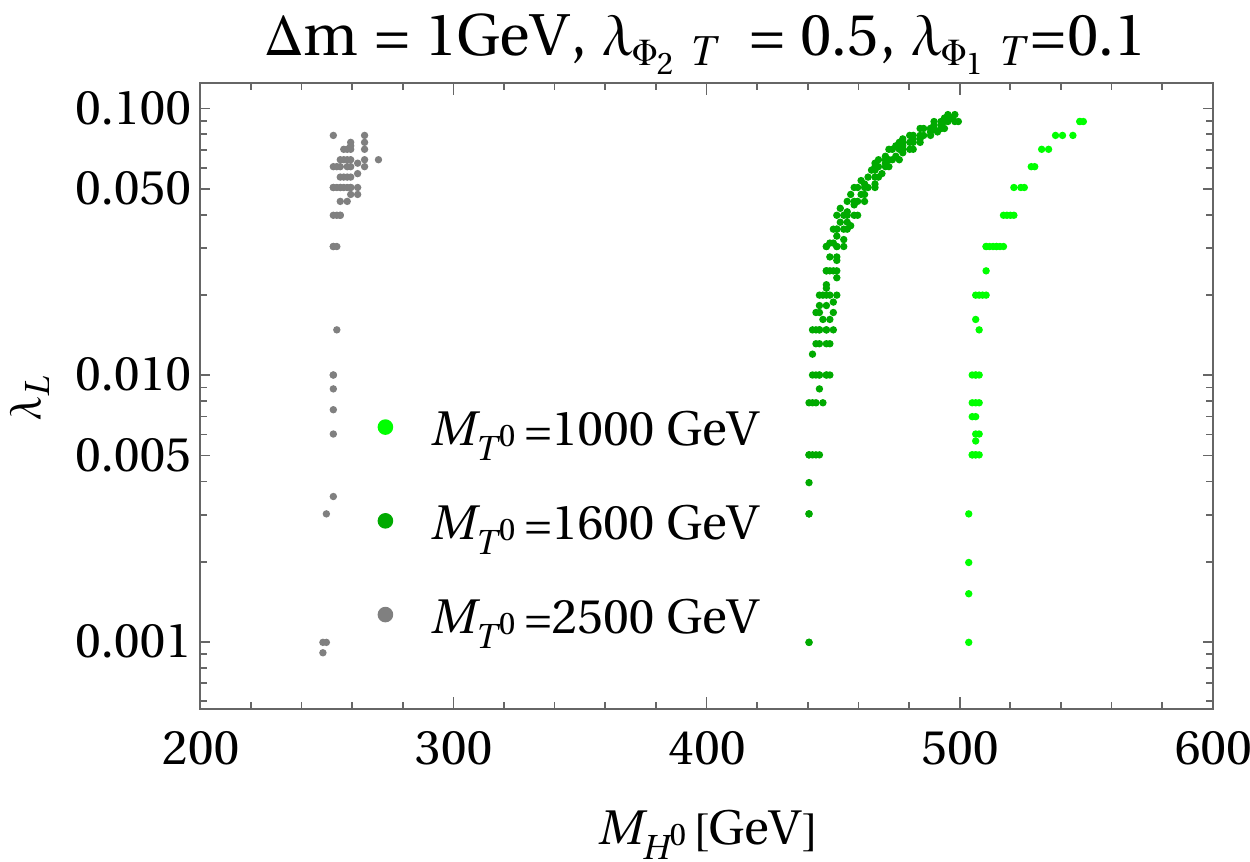}
       \caption{}
      \end{subfigure}
      \begin{subfigure}[b]{0.47\textwidth}
       \includegraphics[scale=0.46]{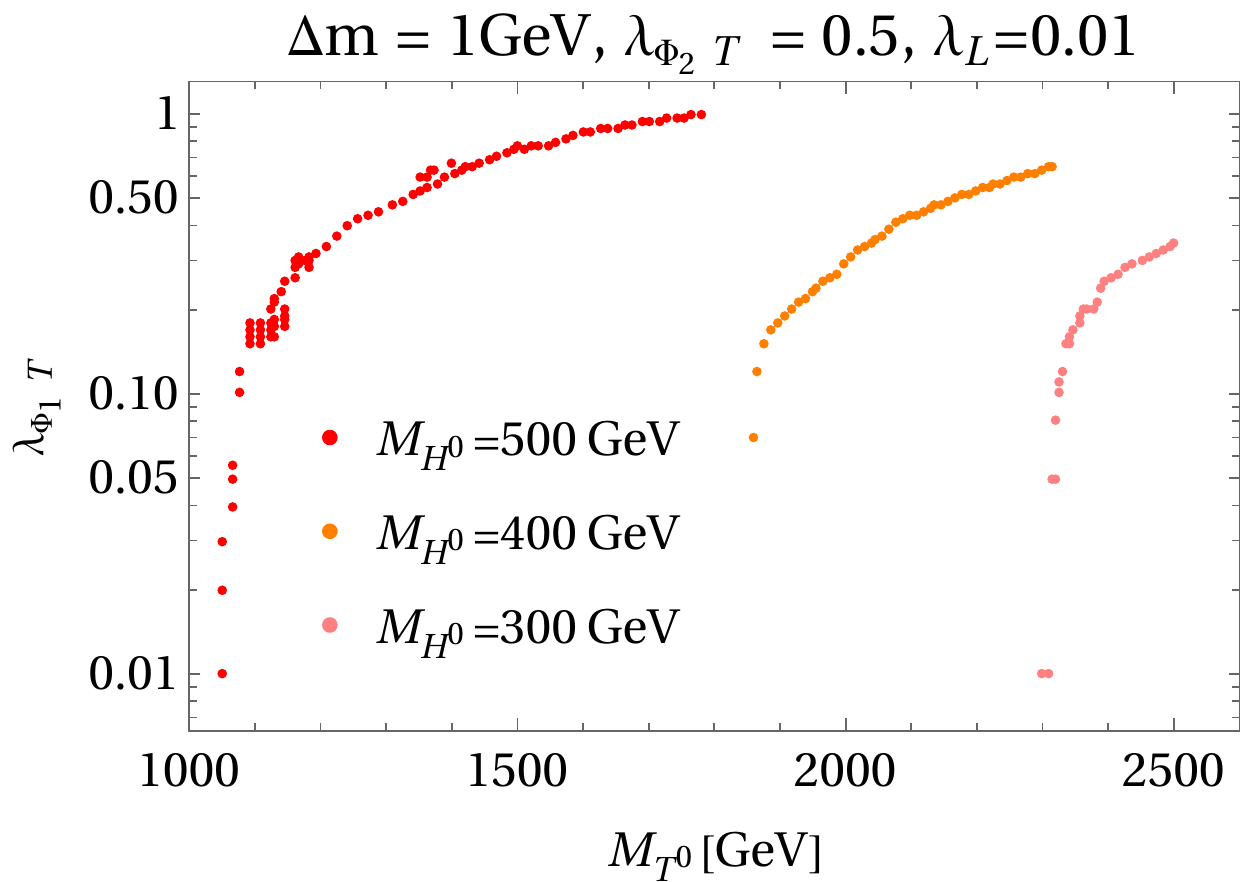}
       \caption{}
      \end{subfigure}
   \caption{Allowed values of the Higgs portal couplings $\lambda_{L}$ and $\lambda_{\Phi_{1}T}$ in terms of the mass of the DM candidates $H^{0}$ and $T^{0}$, respectively. The points satisfy all the theoretical constraints, the correct relic density and direct detection limits.}
    \label{couplingsvsmass}
\end{figure}

In our analysis we kept the values of the Higgs portal couplings $\lambda_{L}$ and $\lambda_{\Phi_{1}T}$ fixed to obtain the results shown in figure~\ref{relicpoints}. However, we also studied the effect of these couplings in satisfying the total relic density and direct detection constraints. In figure~\ref{couplingsvsmass} we show the allowed values of the Higgs portal couplings $\lambda_{L}$ and $\lambda_{\Phi_{1}T}$ that satisfy all theoretical constraints. The points also satisfy the total relic density and direct detection constraints. The left panel shows the behaviour of the $\lambda_{L}$ coupling variation vs. $m_{H^{0}}$ for $m_{T^{0}} = \lbrace{ 1000, 1600, 2500\rbrace}$ GeV. As we increase the value of $\lambda_{L}$, the annihilation rate of $H^{0}$ is increased as well. In order to satisfy the correct relic density, a higher value of the mass $m_{H^{0}}$ is needed. A similar behaviour is seen in the right panel, which shows the variation of the inert triplet-Higgs coupling vs. $m_{T^{0}}$ for $m_{H^{0}} = \lbrace 300, 400, 500 \rbrace$ GeV. To achieve the relic density and direct detection constraints, we need a higher value of the triplet mass as we increase the value of the coupling $\lambda_{\Phi_{1}T}$. Although the variation of the Higgs couplings with each DM candidate leads to changes in satisfying theoretical and experimental constraints, these changes are small. The most important parameter in order to achieve the correct relic density in our model is the DM-DM conversion coupling $\lambda_{\Phi_{2}T}$.

\begin{table}[htb]
\centering
\begin{tabular}{|c|c|c|c|c|c|c|c|c|}
\hline
$M_{H^0}$ & $ M_{T^0}$  & $\lambda_{\Phi_{2}T}$ & $\Omega_{H^{0}}h^{2}$ & $\Omega_{T^0}h^{2}$ & $\Omega_{\text{Tot}} h^{2}$ & $\sigma_{SI}^{H^0} [pb]$ & $\sigma_{SI}^{T^{0}} [pb]$\\
\hline
\hline
170  & 2200 &  0.0 & $1.20\times 10^{-2}$ & $1.08 \times 10^{-1}$ & 0.120 & 5.892 $\times 10^{-11}$ & 8.901 $\times 10^{-12}$ \\
240  & 2500 &  0.5 & $2.35 \times 10^{-2}$ & $9.51 \times 10^{-2}$ & 0.119 & $2.939 \times 10^{-11}$ & 6.830 $\times 10 ^{-12}$ \\
518  & 900  & 0.5 & $1.07 \times 10^{-1}$ & $1.44 \times 10^{-2}$ & 0.121 & 6.852 $\times 10^{-12}$ & 5.693 $\times 10^{-11}$ \\
430  & 2440 & 1.0 & $7.39 \times 10^{-2}$ & $4.74 \times 10^{-2}$ & 0.121 & 9.202 $\times 10^{-12}$ & 7.182 $\times 10 ^{-12}$ \\
506  & 1460 & 1.0 & $1.02 \times 10^{-1}$ & $1.83 \times 10^{-2}$ & 0.120 & 6.909 $\times 10^{-12}$ & 2.083 $\times 10^{-11}$\\
\hline
\end{tabular}
\caption{Benchmarks allowed by positivity, perturbativity, unitarity, oblique parameter corrections, relic density and direct detection constraints. The masses are expressed in GeV.} \label{tb:3}
\end{table}

\section{Searches of Dark Matter in colliders}

\begin{table}[h]
        \centering
        \begin{tabular}{|c|c|c|c|}
          \hline
          $M_{H^0}$[GeV] & $ M_{T^0}$[GeV]  & $\lambda_{\Phi_{2}T}$ & $\Gamma_\text{Tot}$ ($A^{0}$) [GeV] \\
          \hline
          \hline
          170 & 2200 & 0.0 & $7.33\times 10^{-13}$\\
          240 & 2500 & 0.5 & $7.35\times 10^{-13}$\\
          518 & 900  & 0.5 & $7.39\times 10^{-13}$\\
          430 & 2440 & 1.0 & $7.37\times 10^{-13}$\\
          506 & 1460 & 1.0 & $7.38\times 10^{-13}$\\
          \hline
        \end{tabular}
        \caption{Total decay width of $A^{0}$ for benchmarks in table~\ref{tb:3}.}
        \label{tab:my_label}
    \end{table}
    
At colliders, most dark matter searches are driven by the assumption that DM particles are WIMPs. These searches are complementary to those of Direct and Indirect Detection experiments.

In hadron colliders, the main production channels for the IDM correspond to the Drell-Yan production of a $H^{0} A^{0}$ or $H^{0} H^{\pm}$ pair followed by the predominant decay chains $A^{0} \rightarrow Z^{0} H^{0}$ and $H^{0}\rightarrow W^{\pm} h$, leading to a signal of gauge bosons and missing transverse energy. 

The production cross-section at the LHC for the above processes can reach up to 1~pb (3~pb) at 13~TeV (27~TeV)~\cite{Kalinowski2018, Kalinowski2019}.
    
We have calculated the $A^{0}$ total decay width for the benchmarks in table~\ref{tb:3}. The respective values of this width are shown in table~\ref{tab:my_label}. In view of these values, the $A^{0}$ is short-lived enough, therefore we can safely assume that after decoupling they decay promptly mainly into the DM candidate $H^{0}$. 

A real scalar inert triplet could be discovered at the LHC. When the inert triplet $T$ is produced in $pp$ collisions, the charged triplet scalar $T^{\pm}$ decays into the neutral component $T^{0}$ plus a soft-pion or a soft-lepton pair yielding a disappearing charged track in the detector~\cite{Chiang2021}. 

Searches for disappearing tracks at the LHC with 13 TeV data have excluded a real triplet scalar lighter than $248~(275)$~GeV, for a mass splitting of $172 (160)$~MeV with $\textit{L} = 36~{\text{fb}}^{-1}$. 
In both cases, the extra scalar gives rise to the lightest $\mathbb{Z}_2$-odd particle, which does not decay and can contribute as missing energy in the collider~\cite{Wan_2018}.


Due to the $\mathbb{Z}_2$ and $\mathbb{Z}'_2$ symmetries, both inert scalars $H$ and $T$ cannot be coupled to the fermions, and their decays occur only via gauge mode on or off-shell~\cite{Arhrib2014, Belyaev2018}.
The signatures of ITM and IDM are very similar; the only difference is that in the case of the IDM we have an additional neutral pseudoscalar, $A^0$, which can differentiate the signal and thus tell them apart.

\section{Conclusions}\label{Sec:Conclusions}

In this work, we have studied a model with an extended scalar sector that incorporates an inert $SU(2)_L$ doublet, an inert triplet with $Y=0$ and also an active triplet $\Delta$ with $Y=1$. The DM candidates in this model arise from two dark sectors stabilised by the discrete global symmetries $\mathbb{Z}_{2}\times \mathbb{Z}'_{2}$. We have considered a two-component DM scenario in order to reduce the desert regions in the IDM and ITM. 

The relevant parameters for the DM phenomenology are the DM masses $m_{H^{0}}$, $m_{T^{0}}$ and the couplings $\lambda_{L}$, $\lambda_{\Phi_{1}T}$ and $\lambda_{\Phi_{2}T}$. By considering the experimental constraints on the masses of the scalar and pseudoscalar in the IDM, we have focused on the region $100 \leq m_{H^{0}} \leq 550$ GeV for the ID and $100 \leq m_{T^{0}} \leq 2500$ for the IT. Because the $\lambda_{\Phi_{2}T}$ coupling controls the DM-DM conversion, we computed the relic density for the cases $\lambda_{\Phi_2 T}= \{0, 0.5, 1.0\}$ for each DM candidate. For both scalars $H^{0}$ and $T^{0}$, the relic density is under-abundant in each case. However, the contribution of both candidates allow us to obtain the correct total relic density for $M_{H^{0}} \geq 170$ GeV and $M_{T^{0}} \geq 900$ GeV as shown in table~\ref{tb:3} and figure~\ref{relicpoints}. We can reproduce the correct total relic density for $84\%$ of the desert region of $H^{0}$ and about $57\%$ of the desert region of $T^{0}$.
We also have taken into account the upper bounds for the SI DM-nucleon cross-section, where our results for both candidates are within the experimental limits of the PANDAX-II, XENONnT and LUX-ZEPLIN collaborations.

In addition to the DM phenomenology study, we have shown that it is possible to generate light neutrino masses in our model through the type-II seesaw mechanism. A small induced triplet VEV $v_{\Delta} \sim \mathcal{O}(1 $ eV) of the $\Delta$ scalar is required. 
It is worth mentioning that it is possible to generate mater-antimatter asymmetry in the model via leptogenesis, through the CP-violating decays of the active scalar triplet $\Delta$ into leptons $\Delta \to \bar{\ell} \bar{\ell}$~\cite{Datta2022} which will be further studied in a next work.

\section{Acknowledgements}\label{Sec:Acknowledgements}
This work has been founded by PAPIIT project No IN102122. 
JML has been supported by the DGAPA-UNAM Postdoctoral grant. 
SMD thanks the Consejo Nacional de Humanidades Ciencia  y Tecnología (CONAHCYT) Doctorate grant. 
RG and JML thank the Sistema Nacional de Investigadoras e Investigadoras (SNII) of the Consejo Nacional de Humanidades, Ciencia y Tecnología in México (CONAHCYT).

\vspace{0.5cm}
\section*{Data Availability Statement} 
No Data associated in the manuscript.

\bibliography{References.bib}
\end{document}